\documentclass[conference,compsoc]{IEEEtran}
\IEEEoverridecommandlockouts % allow \thanks
\ifCLASSOPTIONcompsoc
  \usepackage[nocompress]{cite}
\else
  \usepackage{cite}
\fi
\ifCLASSINFOpdf
\else
\fi
\usepackage[utf8]{inputenc}
\usepackage[T1]{fontenc}
\usepackage{tikz}
\usepackage{soul}
\usepackage{amsmath}
\usepackage{amssymb}
\usepackage{filecontents}
\usepackage{booktabs}
\usepackage{multirow}
\usepackage{xcolor}
\usepackage{color}
\usepackage{colortbl}
\usepackage{url}
\usepackage{graphicx}
\usepackage{utfsym}
\usepackage{subcaption}
\usepackage{pifont}
\usepackage{bbding}
\usepackage{fontawesome5}
\usepackage{hyperref}
\hypersetup{pdftex,colorlinks=true,allcolors=blue}
\usepackage{array, threeparttable}
\usepackage[]{caption}
\usepackage[linesnumbered, vlined, ruled]{algorithm2e}
\usepackage[misc]{ifsym} % Corresponding letter
\usepackage{fancyhdr}
\definecolor{OliveGreen}{rgb}{0.3,0.7,0.4}
\definecolor{crimson}{RGB}{220,20,60}
\definecolor{ngreen}{RGB}{150,195,125}
\definecolor{nyellow}{RGB}{243,210,102}
\definecolor{nred}{RGB}{216,56,58}
\definecolor{darkred}{rgb}{0.5, 0.0, 0.0}

\usepackage{listings}

\lstdefinelanguage{CapnProto}
{
  morekeywords={struct,union,enum},
  keywordstyle=\color{blue},
  identifierstyle=\color{teal},
  stringstyle=\color{red},
  basicstyle=\tiny\ttfamily,
  commentstyle=\color{green}\ttfamily,
}

\usepackage{tcolorbox}
\newtcolorbox{takeaway}{colframe=black,colback=gray!15,boxrule=1pt,arc=2pt,left=2pt,right=2pt,top=1pt,bottom=1pt, before skip=1em, after skip=0em}

\newif\ifcommentcond
\commentcondtrue % enable comments

\newif\ifupdatecond
\updatecondfalse % disable highlighting
\newcommand{\instruction}[1]{\texttt{#1}}
\newcommand{\xmark}{{\ding{55}}}
\newcommand{\topten}{the top 10}

\newcommand{\topone}{the top 1}

\newcommand{\system}{\textsc{FirmRCA}}
\newcommand{\pomp}{$\mathrm{P}$}
\newcommand{\pompplus}{$\mathrm{P}^+$}

\newcommand{\hookMR}{\texttt{UC\_HOOK\_MEM\_READ\_AFTER}}
\newcommand{\hookMW}{\texttt{UC\_HOOK\_MEM\_WRITE}}
\newcommand{\hookE}{\texttt{UC\_HOOK\_CODE}}

\newcommand{\update}[1]{#1}
\newcommand{\del}[1]{}

\newcounter{yuan} % Yuan
\newcounter{wqy} % wqy
\newcounter{cby} % Boyu
\newcounter{zq} % Zhang Qiao
\newcounter{bin} % Binbin
\newcounter{xu} % Jiacheng
\newcounter{sji} % Prof. Ji
\newcounter{py} % Peiyu

\newcommand{\TotalTestCases}{41}
\newcommand{\TotalPlatforms}{20}
\newcommand{\TotalImages}{17}
\newcommand{\TotalHandlers}{multiple} 
\newcommand{\SystemSuccessNum}{38}
\newcommand{\SystemSuccessRatio}{92.7\%} % SystemSuccessNum / TotalTestCases

\newcommand{\SystemFailedCases}{$C_{12}$, $C_{13}$ and $C_{32}$}

\newcommand{\DeepRCABar}{50\%}
\newcommand{\DeepRCASucess}{10}
\newcommand{\DeepRCATotal}{11}
\newcommand{\DeepRCASample}{$C_{39}$}
\newcommand{\DeepRCADepth}{42,177 (53.6\%)}

\newcommand{\AvgOverheadTimes}{5.47s (5.67$\times$ in addition)}

\newcommand{\AvgOverheadSpace}{23.16 MB}

\newcommand{\TimeBreakDownDepth}{the last 10,000 instructions in the execution trace}
\newcommand{\POMPResolveAvgTime}{1,089.6s} 
\newcommand{\POMPPlusResolveAvgTime}{10.5s} 
\newcommand{\SystemResolveAvgTime}{<0.01s} % 0.004s
\newcommand{\POMPPlusReverseAvgTime}{15.9s}
\newcommand{\POMPReverseAvgTime}{10.2s}
\newcommand{\CompareEffectiveness}{73.2\%} % SystemSuccessRatio(92.7%) - POMP++ 19.5%
\newcommand{\CompareFullTrace}{27.8\%} 

\newcommand{\SystemNoRanking}{\textsc{FirmRCA\_NR}}
\newcommand{\SystemAblationHW}{\textsc{FirmRCA\_HW}}
\newcommand{\SystemAblationRL}{\textsc{FirmRCA\_RL}}

\newcommand{\CompareSummary}{Compared to state-of-the-art works, \system{} demonstrates its superiority in \CompareFullTrace{} improvement in full execution trace analysis capability, polynomial-level acceleration in overall efficiency and \CompareEffectiveness{} higher success rate within \topten{} instructions in effectiveness.}

\begin{document}
% \title{Fuzzing Is Not the End: Efficient Event-based Fault Localization on ARM Embedded Firmware}
\title{\system{}: Towards Post-Fuzzing Analysis on ARM Embedded Firmware with Efficient Event-based Fault Localization}


% compsoc conferences compared to traditional conferences), use this
\author{
\IEEEauthorblockN { 
Boyu Chang\IEEEauthorrefmark{1},
Binbin Zhao\IEEEauthorrefmark{1}\IEEEauthorrefmark{2},
Qiao Zhang\IEEEauthorrefmark{1},
Peiyu Liu\IEEEauthorrefmark{1},
Yuan Tian\IEEEauthorrefmark{3},
Raheem Beyah\IEEEauthorrefmark{2},
Shouling Ji\thanks{Shouling Ji is the corresponding author.}\IEEEauthorrefmark{1}$^{(\textrm{\Letter})}$
}

\IEEEauthorblockA {
  \IEEEauthorrefmark{1}Zhejiang University,
  \IEEEauthorrefmark{2}Georgia Institute of Technology, 
  \IEEEauthorrefmark{3}University of California, Los Angelos \\
}

\IEEEauthorblockA {
  E-mails: bychang@zju.edu.cn, binbin.zhao@gatech.edu, \{j0k1ng, liupeiyu\}@zju.edu.cn, yuant@ucla.edu, \\
  rbeyah@ece.gatech.edu, sji@zju.edu.cn
}
}

% \author{
% \IEEEauthorblockN {

% Boyu Chang\IEEEauthorrefmark{1},
% Binbin Zhao\IEEEauthorrefmark{2},
% Qiao Zhang\IEEEauthorrefmark{1}\IEEEauthorrefmark{3},
% Peiyu Liu\IEEEauthorrefmark{1},
% Yuan Tian\IEEEauthorrefmark{4},
% Raheem Beyah\IEEEauthorrefmark{2},
% Shouling Ji\IEEEauthorrefmark{1}$^{(\textrm{\Letter})}$\thanks{Shouling Ji is the corresponding author.}
% }

% \IEEEauthorblockA {
%   \IEEEauthorrefmark{1}Zhejiang University,
%   \IEEEauthorrefmark{2}Georgia Institute of Technology, \\
%   \IEEEauthorrefmark{3}Hong Kong University of Science and Technology,
%   \IEEEauthorrefmark{4}University of California, Los Angelos \\
% }

% \IEEEauthorblockA {
%   E-mails: bychang@zju.edu.cn, binbin.zhao@gatech.edu, qzhangdi@connect.ust.hk, liupeiyu@zju.edu.cn, yuant@ucla.edu, rbeyah@ece.gatech.edu, sji@zju.edu.cn
% }
% }


\maketitle

\begin{abstract}

    While fuzzing has demonstrated its effectiveness in exposing vulnerabilities within embedded firmware, the discovery of crashing test cases is only the first step in improving the security of these critical systems.
    The subsequent fault localization process, which aims to precisely identify the root causes of observed crashes, is a crucial yet time-consuming post-fuzzing work.
    Unfortunately, the automated root cause analysis on embedded firmware crashes remains an underexplored area, which is challenging from several perspectives:
    (1) the fuzzing campaign towards the embedded firmware lacks adequate debugging mechanisms, making it hard to automatically extract essential runtime information for analysis;
    (2) the inherent raw binary nature of embedded firmware often leads to over-tainted and noisy suspicious instructions, which provides limited guidance for analysts in manually investigating the root cause and remediating the underlying vulnerability.

    To address these challenges, we design and implement \system{}, a practical fault localization framework tailored specifically for embedded firmware. 
    \system{} introduces an event-based footprint collection approach that leverages concrete memory accesses in the crash reproducing process to aid and significantly expedite reverse execution. 
    Next, to solve the complicated memory alias problem, \system{} proposes a history-driven method by tracking data propagation through the execution trace, enabling precise identification of deep crash origins.
    Finally, \system{} proposes a novel strategy to highlight key instructions related to the root cause, providing practical guidance in the final investigation.
    To demonstrate the efficacy of \system{}, we evaluate it with both synthetic and real-world targets, including \TotalTestCases{} crashing test cases across \TotalImages{} firmware images.
    The results show that \system{} can effectively (\SystemSuccessRatio{} success rate) identify the root cause of crashing test cases within \topten{} instructions.
    \CompareSummary{}
\end{abstract}

% \fi
% \IEEEpeerreviewmaketitle

\section{Introduction}

Firmware is a specialized software program embedded in hardware devices, providing essential functionalities for initial configurations, managing communications, and performing I/O operations \cite{wright2021challenges}.
It plays a pivotal role in our interconnected world, serving as a critical component of devices employed across various infrastructure sectors, such as advanced manufacturing and intelligent healthcare.
Given the ubiquity of firmware and its critical role in the proper functioning of these systems, vulnerabilities in firmware can lead to severe consequences, potentially allowing malicious attackers to install malware \cite{hpreport}, steal sensitive data \cite{ibm-databreach}, and even remotely control compromised devices \cite{viasat}.

With the growing risk of firmware attacks on embedded devices, considerable efforts are needed to mitigate firmware vulnerabilities.
One of the most effective methods for discovering these faults is fuzzing, which involves providing randomly generated test cases to explore program behaviors  \cite{scharnowski2022fuzzware,li2022muafl,zaddach2014avatar,muench2018avatar}.
When the firmware crashes, the fuzzer saves the corresponding test case for later analysis.
However, discovering crashing test cases is only the first step in improving firmware security.
Identifying the root causes of firmware crashes is a crucial step after fuzzing, as it provides a deeper understanding of the underlying vulnerabilities.

Given crashing test cases, analysts have to undertake the complex task of manually analyzing their root causes, a process known as \textit{fault localization}. This task is extremely tedious and prone to human error \cite{xu2017postmortem,mu2022depth}.
Over the past few decades, researchers have proposed various automated fault localization techniques.
\textit{Spectrum-based Fault Localization} methods involve mutating the initial crashing test case to collect two large sets of crashing and non-crashing test cases with similar execution traces. They then statistically correlate program entities (e.g., statements and instructions) with the observed crash \cite{blazytko2020aurora,park2023benzene,xuracing}.
\textit{Postmortem-based Fault Localization} methods start with debugging files (e.g., core dumps, execution traces, and memory snapshots), then perform reverse execution and backward taint analysis to track the propagation of invalid data \cite{guo2019deepvsa,cui2018rept,cui2016retracer}.
More recently, \textit{Learning-based Fault Localization} approaches have utilized neural networks to understand the code context, then highlight statements or instructions related to the crash \cite{li2017transforming,sohn2017fluccs,li2019deepfl,yang2024large}.

Despite the variety of techniques proposed for software running on resource-rich systems with abundant debugging support, automated firmware crash analysis remains an underexplored area.
Identifying the root cause of a firmware crash is still non-trivial due to the following challenges. 

\noindent\textbf{Challenge \uppercase\expandafter{\romannumeral1}: Inadequate debugging mechanisms.}
Intuitively, analysts require a comprehensive understanding of the crash to figure out its root cause.
Runtime information, such as the execution trace, register and memory values during the execution leading up to the crash, can provide valuable insights for analysts.
Unfortunately, fuzzing campaigns towards the embedded firmware lack debugging mechanisms (e.g., sanitizers) to automatically identify and extract key information related to the crash.
Analysts have to dive into tens of thousands of instructions only with the support of a basic program debugger or raw logs \cite{scharnowski2022fuzzware} to manually examine both control flow and data flow, which is extremely tedious, time-consuming, and prone to human mistakes.
Considering the lengthy execution trace and rich runtime information, it is challenging to automatically extract and utilize essential runtime information for firmware fault localization.

\noindent\textbf{Challenge \uppercase\expandafter{\romannumeral2}: Limited investigation guidance.}
The embedded firmware is usually stripped into a raw binary file where debugging symbols (e.g., function and variable names) are stripped away, leaving only a bunch of instructions mixed with data.
The lack of semantic information in the stripped firmware binary overwhelms the analysts with an abundance of potentially suspicious instructions that still require non-trivial manual investigation.
However, previous works either treat all instructions uniformly \cite{xu2017postmortem,cui2018rept} or fail to differentiate instructions within the same basic block by design \cite{wong2012software,sarhan2022survey}.
These over-tainted and noisy instructions offer limited practical guidance for the final manual root cause investigation.

Apart from the aforementioned two challenges, there is another common and crucial challenge towards fault localization.
Given the execution trace leading up to a crash, analysts can reversely analyze the instructions to figure out how a pointer is initialized with corrupted data, propagated through function calls, and ultimately dereferenced illegally.
However, it is tedious and time-consuming to manually dig out the root cause through tens of thousands of instructions.
For the automated root cause analysis, the primary obstacle is the memory alias problem.
Memory aliasing, where multiple pointers may point to the same underlying memory location, introduces uncertainty into the analysis of the data flow of the crashing test case and greatly complicates the task of accurately tracking data propagation \cite{cui2018rept, cui2016retracer}. %% GPT

\noindent\textbf{Our solution.}
In light of these challenges, we propose \system{}, a practical postmortem-based fault localization framework.
\system{} aims to identify faulty instructions of a crash reported by the embedded firmware fuzzer.
To solve the \textbf{Challenge \uppercase\expandafter{\romannumeral1}}, we design an event-based footprint collection method on top of firmware rehosting to record essential runtime information.
Our key intuition is that \textit{concrete memory accesses} in the crash reproducing process can aid the reverse execution phase and accelerate resolving memory aliases significantly in fault localization.
One concrete memory access includes the address of the instruction, the access method (i.e., read or write), the source (destination) register, and the concrete destination (source) memory address.
To tackle the \textbf{Challenge \uppercase\expandafter{\romannumeral2}}, we design a novel two-phase root cause analysis method, which dynamically computes suspicious scores for all instructions.
Based on our key insight that not all tainted instructions share the same contribution to the root cause, we further propose two ranking strategies to highlight the instructions that have higher priorities for analysts to investigate.
For the common challenge of the memory alias problem, we restore concrete memory accesses from the collected event-based footprint.

In summary, we make the following contributions.

$\bullet$ We propose \system{}, a practical postmortem-based automated fault localization system tailored specifically for addressing the unique challenges of embedded firmware. 
We introduce an event-based approach for \system{} to achieve efficient instruction analysis and provide accurate root causes even for extremely long execution traces.
Our event-based approach utilizes concrete memory accesses to resolve the memory alias problem and greatly improves overall effectiveness and efficiency.

$\bullet$ We propose a novel strategy to analyze the behavior of instructions and then assign suspicious scores to prioritize attention toward more critical instructions.
Our strategy provides practical guidance for analysts in identifying the root causes that are far from the crash site.

$\bullet$ We have implemented and evaluated \system{} with \TotalTestCases{} crashing test cases across \TotalImages{} firmware images.
The results show that \system{} successfully identifies \SystemSuccessNum{} out of \TotalTestCases{} crashing test cases within \topten{} instructions with a success rate of \SystemSuccessRatio{}.
\CompareSummary{}
We open source \system{} at \url{https://github.com/NESA-Lab/FirmRCA}.

\section{Background}
\vspace{-4pt}

\noindent\textbf{Firmware rehosting.} 
In firmware analysis, researchers propose firmware rehosting due to the slow nature and high costs of using actual hardware \cite{fasano2021sok}.
Firmware rehosting is a technique to execute firmware binaries in an emulated environment instead of their original hardware devices.
The emulation process replaces the need for the physical hardware, allowing users to analyze and even manipulate firmware code in a controlled virtual setting.
Due to the effectiveness of firmware rehosting, recent works have adopted this technique for vulnerability discovery \cite{scharnowski2022fuzzware,scharnowski2023hoedur,chesser2023icicle}.

In the context of firmware rehosting, the hook function is a powerful method for intercepting specific actions.
For instance, when using Unicorn \cite{unicorn} to emulate a firmware binary, one can design and implement a \texttt{simple\_test} function, and register it with the \hookE{} option.
When executing the binary, the \texttt{simple\_test} function is called before the program executes any instructions.
Although hooks provide analysts with the ability to monitor the execution process, it is still hard to collect key runtime information related to the crash for the sake of both effectiveness and efficiency in fault localization.
In this work, we propose an event-based footprint collection method on top of firmware rehosting to facilitate fault localization.

\noindent\textbf{Reverse execution.}
In the field of postmortem-based fault localization, analysts perform reverse execution to gain deeper insights into program crashes \cite{cui2018rept, cui2016retracer}.
The key idea of reverse execution is to \textit{undo} the executed instruction to revert registers and memory values to their previous states, providing analysts with a unique perspective on the evolution of program state over time.
To undo the executed instruction, analysts usually design complicated handlers that inversely execute the instruction according to its behavior (i.e., inverse handlers).
Besides, the process of reverse execution heavily relies on runtime data collected from the system, typically including detailed logs, memory dumps, and other relevant data that encapsulate the state of the system. 
In this work, we utilize our collected footprint to improve reverse execution and design multiple inverse handlers for the 32-bit ARM Cortex-M embedded firmware.

\noindent\textbf{Use-define chain.}
In the field of data flow analysis, a Use-Define Chain (UD Chain) is a data structure that describes a use and all the definitions of a variable in the program, where that variable can reach the use without any other intervening definitions \cite{alfred2007compilers}.
The use-define chain consists of a bunch of use nodes and define nodes, providing a way to track the data flow by establishing a connection between uses and definitions of different variables.
In the firmware analysis scenario, constructing an accurate use-define chain is non-trivial due to the closed-source and symbol-stripped nature of firmware.
In this work, we propose an automated method that utilizes the execution trace and disassembled firmware binary to improve the effectiveness of use-define chain construction, upon which we further analyze the propagation of invalid data and taint relevant instructions.

\section{Overview}
\vspace{-4pt}

In this section, we first describe the problem scope and the goal of \system{}.
We then present a high-level overview of \system{}'s architecture and its components.

\vspace{-8pt}
\subsection{Problem Scope}\label{threat model}
\vspace{-4pt}

\system{} is focused on automating the identification of errors in embedded firmware, representing a step forward in addressing firmware vulnerabilities after fuzzing activities.
In this scenario, analysts have completed the rehosting-based fuzzing campaign towards the raw binary firmware and then start to analyze crashes\cite{scharnowski2022fuzzware}.
Given a large set of crashing test cases generated in the fuzzing process, our goal is to efficiently and effectively identify instructions pertaining to the crash for every corresponding crashing test case.
\system{} highlights and ranks suspicious instructions, providing a starting point for the manual investigation to aid both novice and expert analysts.

\begin{figure}[htbp]
    \centering
    \begin{subfigure}{0.45\textwidth}
        \includegraphics[width=\textwidth]{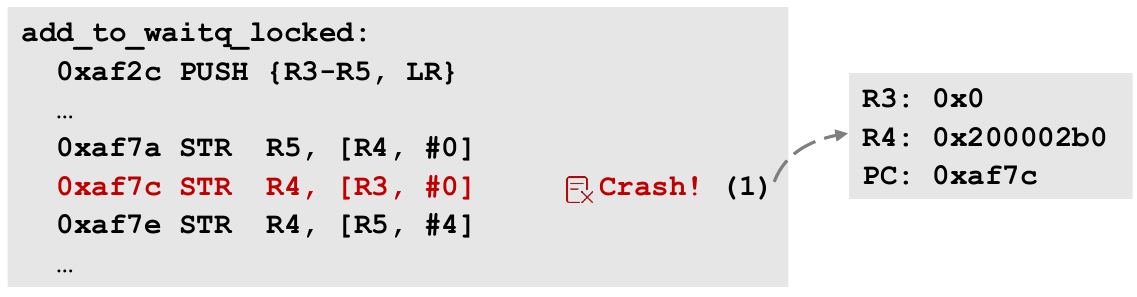}
        \caption{Invalid memory write}
        \label{fig:crash-type-a}
    \end{subfigure}
    \hfill
    \begin{subfigure}{0.45\textwidth}
        \includegraphics[width=\textwidth]{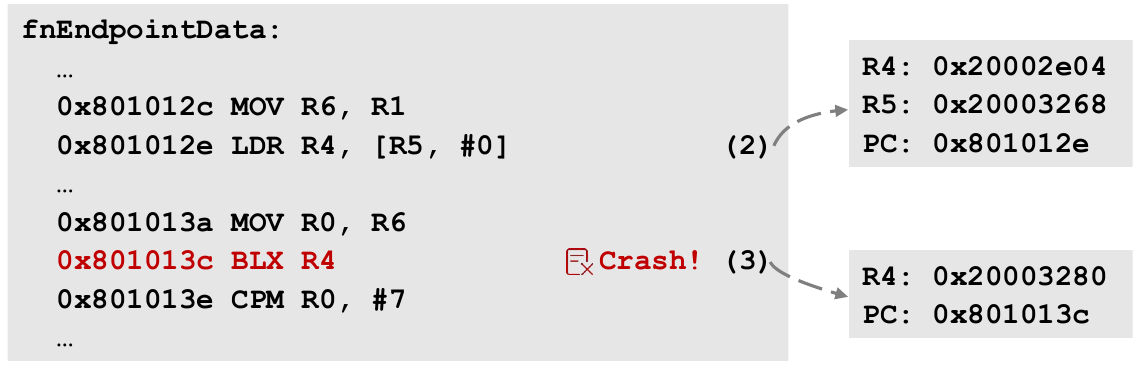}
        \caption{Invalid instruction execution}
        \label{fig:crash-type-b}
    \end{subfigure}
    \caption{Two instruction snippets of direct reasons to trigger firmware crashes. The relevant register values before the execution of the instruction are shown on the right. The snippet (a) shows an invalid memory write and the snippet (b) shows an invalid instruction execution.}
    \label{fig:crash-type} 
\end{figure}

The root causes of real-world crashes are diverse and complicated (e.g., use after free, double free, and out-of-bounds write).
It is not scalable to model the crashes for each crash type to identify their root causes.
However, at the instruction level, the majority of firmware crashes can be attributed to two main reasons, including invalid memory accesses and invalid instruction executions \cite{xu2017postmortem}.
In this work, we call these two reasons as direct reasons for crashes.
Figure~\ref{fig:crash-type} shows two instruction snippets of direct reasons.
The relevant register values before executing the instruction are detailed on the right.
Figure~\ref{fig:crash-type-a} shows an instruction snippet of the crash site for CVE-2021-3329.
When executing \instruction{STR R4, [R3, \#0]} at \texttt{(1)}, the program tries to store the value of \instruction{R4} (i.e., \instruction{0x200002b0}) in the memory at the address \instruction{R3 + 0}.
However, the register \instruction{R3} happens to be zero, which triggers an unmapped memory access at \instruction{R3 + 0} (i.e., zero).
Then, the program crashes because of an invalid memory write.
Figure~\ref{fig:crash-type-b} illustrates another crash reason.
When the \instruction{BLX R4} instruction is executed, the program attempts to branch to the address of \instruction{R4} (i.e., \instruction{0x20003280}), which is a function pointer in the memory area.
However, the function pointer references an invalid instruction (i.e., a NULL pointer) at runtime, resulting in a crash.
While these direct reasons may not always be the ultimate root causes in all cases, they provide \system{} with the key information (i.e., \instruction{R3} in Figure~\ref{fig:crash-type-a} and \instruction{R4} in Figure~\ref{fig:crash-type-b}) to perform fault localization, serving as a first step in revealing the underlying issues. 
Accordingly, instead of modeling tremendous crash types, the \system{} framework is designed to handle these two direct reasons to achieve the best scalability.

\vspace{-0.1cm}
\subsection{Architecture}
\vspace{-0.1cm}

\begin{figure*}[htbp]
    \centering
    \setlength{\belowcaptionskip}{-2mm}
    \includegraphics[width=0.95\textwidth]{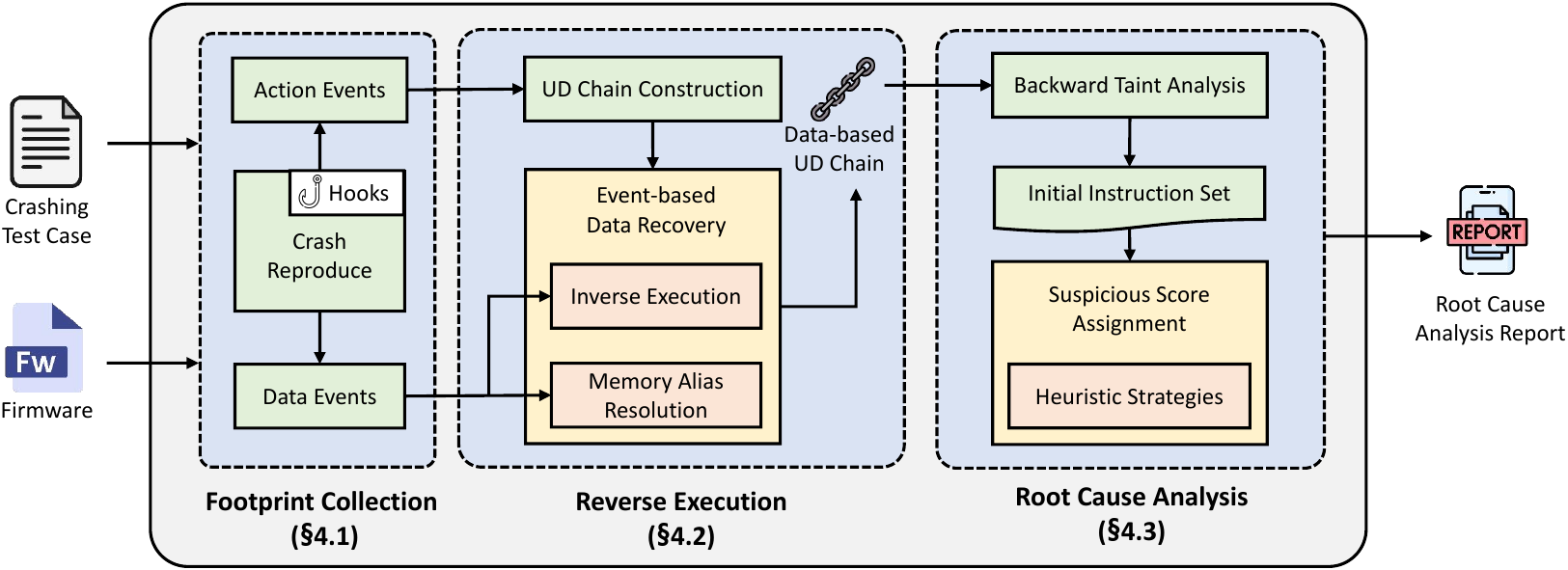} 
    \caption{Overview of \system{}. The architecture of \system{} consists of three components, including footprint collection, reverse execution, and root cause analysis.}
    \label{fig:Overview} 
\end{figure*}

Figure~\ref{fig:Overview} illustrates the high-level architecture of \system{}, which consists of three key components: footprint collection, reverse execution, and root cause analysis. 

\noindent\textbf{Footprint collection.}
To enable effective analysis of firmware crashes, \system{} first executes the firmware binary with the crashing test case and collects runtime data to support the subsequent reverse execution and root cause analysis.
To address the challenge of inadequate debugging mechanisms, \system{} proposes an event-based footprint collection approach.

\noindent\textbf{Reverse execution.}
The main purpose of reverse execution is to construct the dynamic propagation of data in a crashing test case.
To tackle the prevalent issue of memory aliases that can obscure data flow analysis, \system{} employs a history-driven reverse execution method. 
By leveraging the concrete memory accesses captured during the footprint collection phase, \system{} efficiently and accurately resolves memory aliases, enabling more precise tracking of data dependencies. 

\noindent\textbf{Root cause analysis.}
As aforementioned in Section~\ref{threat model}, \system{} targets crashes triggered by an invalid memory access or an invalid instruction execution.
In both cases, \system{} treats the corresponding register and memory address as the taint sink. 
\system{} then performs a novel two-phase root cause analysis.
First, \system{} performs common backward taint analysis to obtain an initial instruction set.

Next, to provide instructive guidance for further investigation, \system{} then ranks the tainted instructions according to their behaviors, highlighting the most likely contributors to the root cause. 
\section{Design}

In this section, we introduce the design of \system{}.
The main workflow is as follows.
First, \system{} accepts a crashing test case generated by fuzzers and the corresponding firmware binary as inputs.
Second, in the footprint collection phase, \system{} reproduces the crash to collect runtime data, including instruction addresses, register and memory values.
Third, the reverse execution component uses action events to build a use-define chain and data events to recover unknown data to further construct a data-based use-define chain.
Then, \system{} carries out backward taint analysis along the use-define chain to identify instructions related to the crash.
Finally, \system{} adopts heuristic ranking strategies to highlight key instructions as the root cause for analysts to investigate.

\subsection{Footprint Collection}\label{sec:footprint-collection}
To effectively analyze the root causes of firmware crashes, it is crucial for \system{} to gather essential runtime information about the crashing test cases.
We design an event-based footprint collection methodology on top of firmware rehosting.
Specifically, \system{} identifies two types of key events for understanding firmware crashes, including data events and action events.
In the following, we introduce details of the event-based footprint collection methodology of \system{}.

\begin{figure}[tbp]
    \centering
    \setlength{\belowcaptionskip}{-3mm}
    \includegraphics[width=0.45\textwidth]{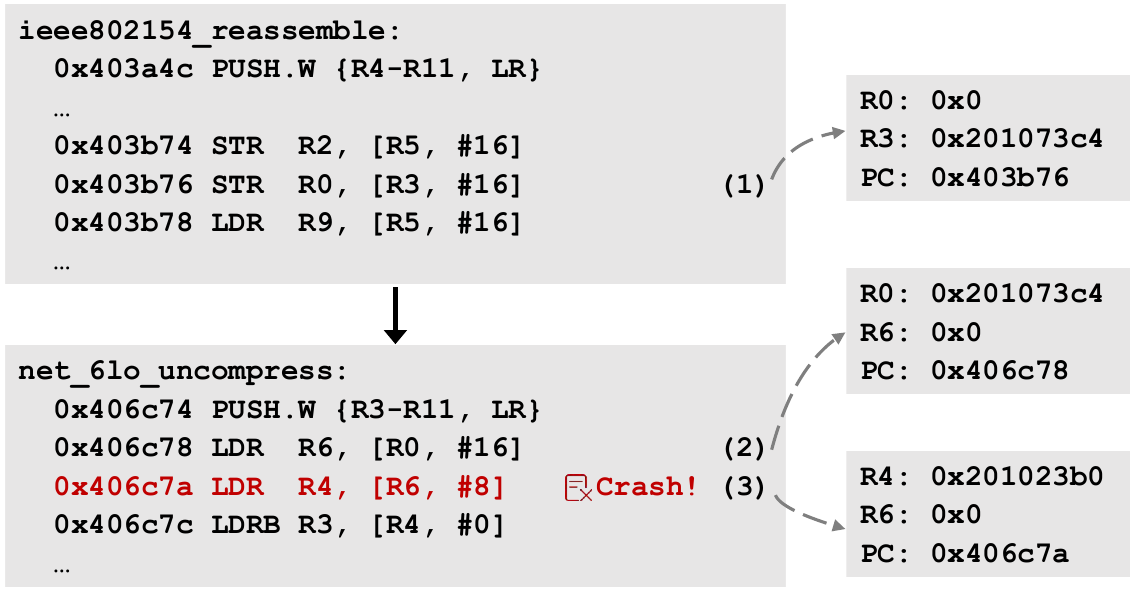} 
    \caption{An instruction snippet of CVE-2021-3322, where a radio frame is inserted and marked for fragmentation. The frame contains the full payload size, leading to an unexpected pointer alias, which finally results in an unexpected NULL pointer dereference in \texttt{net\_6lo\_uncompress} function.}
    \label{fig:null-ptr-example} 
\end{figure}

\subsubsection{Data Events}

Many firmware crashes are triggered by unexpected memory accesses. 
Figure~\ref{fig:null-ptr-example} shows an example of a crash due to an invalid read.
The program first stores \instruction{R0} in the memory at the address \instruction{R3 + 16} at \texttt{(1)} in \instruction{ieee802154\_reassemble} function, then loads value from the address of \instruction{R0 + 16} and assigns it to \instruction{R6} at \texttt{(2)} in \instruction{net\_6lo\_uncompress} function.
Finally, the program loads value from the address of \instruction{R6 + 8} which results in a NULL pointer dereference at \texttt{(3)}.

However, given a crashing test case and the firmware binary, it is non-trivial to obtain such knowledge to understand the crash.
Analysts have to dive into runtime details, examine every instruction from the crash site and finally figure out the fact that it is \instruction{R0 + 16} and \instruction{R3 + 16} pointing to the same address (i.e., \instruction{0x201073d4}) together with an invalid value (i.e., zero) of \instruction{R0} at \texttt{(1)} that induces the crash.
As there are a large number of instructions that are executed between \texttt{(1)} and \texttt{(2)}, it is time-consuming and error-prone to manually analyze the crash reason.
In addition to manual analysis, the effectiveness of automated analysis methods is also limited by a large number of instructions.
Existing automated analysis methods rely on the correct recovery of registers at the crash site and memory data from crash reports or core dump files \cite{cui2018rept,mu2019pomp++,xu2017postmortem,cui2016retracer}.
However, these methods can only recover a part of data values along the execution trace due to inherent analysis limitations (e.g., memory aliases and irreversible instructions).
The longer the execution trace to analyze, the more data can not be recovered and the less accurate the result will be.
Therefore, it is difficult to pinpoint the exact root cause.

To overcome this difficulty, we propose a history-driven method that leverages concrete memory accesses to reconstruct the data flow for the crashing test case.
\system{} views the memory access behavior as a data event.
Memory access is either a read or write, thus \system{} pays attention to two types of data events.
One data event is a successful memory read and the other is a successful memory write.
\system{} monitors these two events and immediately pauses emulation as soon as either event occurs.
Then, \system{} saves the details of the data event.
We design a data structure to store the type of memory access (i.e. read or write), the address of the instruction being executed,the actual data being read (written), and the corresponding memory address. 
Based on data events, \system{} obtains concrete memory access information and efficiently understands the data flow between memory reads and writes even for extremely long execution traces.

\subsubsection{Action Events}

In addition to the memory access history captured as data events, the execution trace is another crucial piece of information that is utilized by \system{} to enable effective fault localization. 
We consider the execution trace as a sequence of instructions executed by the firmware. 
Through the analysis of the execution trace, analysts can understand the behavior of the program leading up to the crash, thereby focusing their investigation on the root cause more effectively. 

\begin{figure}[htbp]
    \centering
    \setlength{\belowcaptionskip}{-2mm}
    \includegraphics[width=0.45\textwidth]{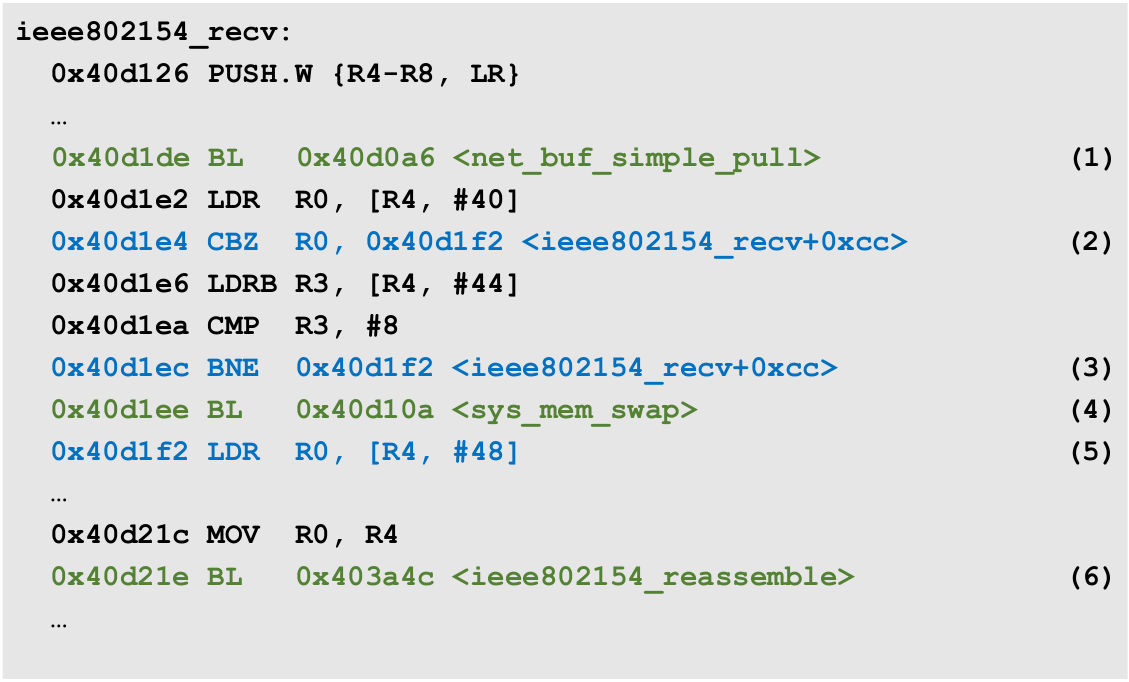} 
    \caption{An instruction snippet of the firmware for CVE-2021-3322. Inconsecutive instructions for \update{intraprocedural and interprocedural} analysis are highlighted in different colors.}
    \label{fig:intra-and-inter} 
\end{figure}

Nevertheless, it is difficult to obtain the complete execution trace of a crashing test case.
A straightforward approach is to iteratively look for the previous instruction from the crash site, and track \instruction{pc} values stashed on the stack for function calls.
The main challenge here is limited analysis depth caused by inconsecutive instructions.
Figure~\ref{fig:intra-and-inter} illustrates two main kinds of inconsecutive instructions in \update{intraprocedural and interprocedural} analysis.
For \update{intraprocedural} analysis, the key issue is the inability to reliably resolve conditional branches without access to the actual runtime register values.
Although we can construct a control flow graph and find all preceding blocks of any instruction with static forward analysis, it is still challenging to determine the correct block because of the lack of accurate register values for conditional branches.
Consider the example instruction \instruction{LDR R0, [R4, \#48]} at \texttt{(5)}: while static analysis can identify the two preceding instructions at \texttt{(2)} and \texttt{(3)}, the correct preceding block ultimately depends on the values of registers \instruction{R0} and \instruction{R3}, which are unknown. 
Incorporating these implicit register dependencies into the analysis can introduce prohibitive overhead, rendering the approach impractical. 
Similarly, for \update{interprocedural} analysis, the challenge lies in the lack of visibility into sibling function executions. 
Even if the call chain can be reconstructed from the stack trace, there is no way to know whether a relevant function has been invoked or not, as its stack frame would have been popped off. 
In the example of the \instruction{BL 0x403a4c} at \texttt{(6)}, the stack frames for \instruction{net\_buf\_simple\_pull} at \texttt{(1)} and \instruction{sys\_mem\_swap} at \texttt{(4)} have already been removed, potentially obscuring crucial context for understanding the root cause.

To address this challenge, we hold the idea of breaking the whole into parts and propose a trigger-based method to first collect each execution of one instruction and then recover the complete execution trace. 
More specifically, \system{} first regards the execution of one instruction as an \textit{action event}, which is triggered when an instruction is about to be executed.
In the ARM architecture, register \instruction{PC} saves the address of the next instruction to execute.
\system{} pauses the execution when an action event is triggered and then records the value of \instruction{PC}.
Finally, \system{} successfully recovers the complete execution trace by combining the collected action events, which helps identify the root cause that is far from the crash site.

\begin{figure}[htbp]
    \centering
    \setlength{\belowcaptionskip}{-2mm}
    \includegraphics[width=0.4\textwidth]{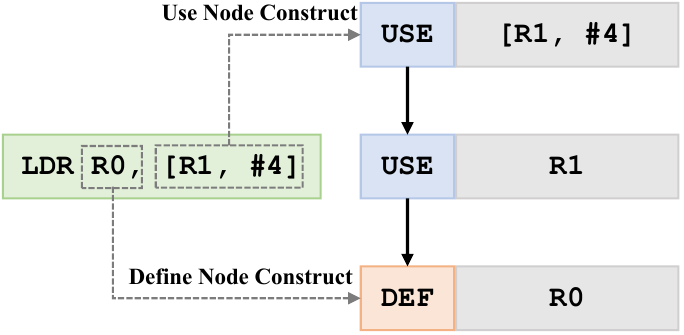} 
    \caption{An example of constructing the use-define chain for the instruction \instruction{LDR R0, [R1, \#4]}. The source operand \instruction{[R1, \#4]} and its base register \instruction{R1} generate two use nodes. The destination operand \instruction{R0} generates a define node.}
    \label{fig:ud-chain} 
\end{figure}

\subsection{Reverse Execution}\label{sec:reverse-execution}

Through the event-based footprint collection phase, \system{} obtains runtime data consisting of memory accesses and executed instruction addresses.
In this section, we introduce how \system{} utilizes this information to perform the reverse execution for further understanding the propagation of the corrupted data.

For an invalid memory access or an invalid instruction execution that triggers a crash, the basic idea for identifying the root cause is to reversely examine instructions in the execution trace from the crash site and figure out why the corresponding data has been corrupted, which is called reverse execution.
Unfortunately, even with a complete execution trace available, determining the precise data propagation and correlating individual instructions is a non-trivial challenge, primarily due to the memory alias problem.
Consider again the example in Figure~\ref{fig:null-ptr-example}, where the instructions at \texttt{(1)} and \texttt{(2)} access the same memory address (i.e., \instruction{0x201073d4}).
If we could establish this correlation, we could then blame the instruction at \texttt{(1)} as the root cause of the crash.
However, the main difficulty is that the concrete memory addresses involved in these instructions are often obscured by the presence of irreversible instructions.
Take the instruction \instruction{MOV R3, \#0} that sets the register \instruction{R3} to zero for example.
We only know that \instruction{R3} holds zero after execution, and there is no hint to recover the original value of \instruction{R3} prior to the execution of this instruction.
In this case, the reverse execution fails to handle other instructions that use the old value of \instruction{R3} before the instruction \instruction{MOV R3, \#0}.
To tackle this challenge, \system{} adopts a two-step history-driven method to perform the reverse execution.
First, we construct a use-define chain for analyzing execution traces.
Then, on top of the use-define chain, we inversely execute the instruction to recover registers into previous states.
We detail our design as follows.

\subsubsection{Use-define Chain Construction}

The first step in \system{}'s reverse execution process is to iteratively reconstruct the complete execution trace using the action events collected during the footprint collection phase. 
For each instruction encountered, \system{} carefully parses the instruction's behavior, extracting the source and destination operands.
Figure~\ref{fig:ud-chain} illustrates this use-define chain construction process.
Taking \instruction{LDR R0, [R1, \#4]} as an example.
In this case, the firmware reads four bytes from the memory address of \instruction{R1 + 4} and stores them in \instruction{R0}.
The operand \instruction{[R1, \#4]} is considered a source operand, and \instruction{R0} is considered a destination operand.
These source and destination operands represent the data flow within the instruction.
To obtain the data flow between instructions, \system{} further utilizes the two operands to generate \textit{use nodes} and \textit{define nodes}. 
Specifically, a use node describes a read behavior for a register or memory address, while a define node represents a write behavior.
For the \instruction{LDR R0, [R1, \#4]} instruction, the destination operand \instruction{R0} is used to create a define node, and the source operand \instruction{[R1, \#4]} is used to create a use node.
Additionally, \system{} also creates extra use nodes for the base register and index register involved in the memory operand.
It is important to note that a single instruction may generate multiple explicit and implicit use and define nodes, as \system{} thoroughly analyzes the semantics of each instruction to identify all the relevant source and destination operands. 
As \system{} parses each new instruction in the execution trace, the corresponding use and define nodes are appended to the growing use-define chain.
This carefully constructed use-define chain serves as the foundation for the subsequent data recovery step in \system{}'s reverse execution process.

\subsubsection{Event-based Data Recovery}

After constructing the use-define chain, the next step in \system{}'s reverse execution process is to inversely recover the register values preceding each instruction's execution. 
To achieve this, \system{} employs custom handlers designed to handle specific types of instructions.
For instance, we design \instruction{add\_handler} for the add operation.
From the instruction \instruction{add R0, R1}, we derive \instruction{R0' = R0 + R1}, \instruction{R0 = R0' - R1} and \instruction{R1 = R0' - R0} where \instruction{R0'} denotes the value of register \instruction{R0} after execution.
If there is only one unknown value in three operands, we can recover it using the other two values and update the corresponding node.
Whenever a node in the use-define chain is updated, \system{} examines the entire chain for other possible updates that can be propagated. 
This iterative process continues until the use-define chain converges, at which point \system{} has recovered as many previous states as possible.
However, these methods are insufficient to process the node with memory alias, as aforementioned in Figure~\ref{fig:null-ptr-example}.
To solve this problem, \system{} utilizes data events to recover the actual runtime data.
For example, consider the \instruction{STR R0, [R1, \#4]} instruction that stores the value of \instruction{R0} in the memory at the address \instruction{R1 + 4}.
During the previous footprint collection phase, \system{} has obtained the target address (i.e., \instruction{[R1, \#4]}) and concrete data to be stored.
Armed with this data event, \system{} can directly recover the value of \instruction{R0} from the concrete data, and the value of \instruction{R1} by subtracting 4 bytes from the target address.
Compared to other techniques such as hypothesis testing \cite{xu2017postmortem} and value-set analysis \cite{mu2019pomp++}, the data event not only guarantees accuracy, but also greatly accelerates reverse execution.

\subsection{Root Cause Analysis}\label{sec:root-cause-analysis}

In this section, we introduce a novel two-phase root cause analysis method.
First, \system{}  performs common backward taint analysis to obtain an initial instruction set, where the instructions are associated with the root cause.
Then, \system{} assigns suspicious scores for all tainted instructions to rank and highlight the instructions that have higher priorities in further investigation.

\subsubsection{Backward Taint Analysis}

The backward taint analysis starts by identifying the direct reason for the crash. 
\system{} then leverages this information to track the propagation of invalid data backward through the execution trace.
In the following, we introduce how \system{} sets the taint sink and taints instructions.

\noindent\textbf{Taint sink.}
The taint sink is the starting point of backward taint analysis.
\system{} identifies concrete memory address or registers as the taint sink according to the instruction type at the crash site.
For invalid memory read (write) instructions, both the base and index registers are marked as taint sinks.
As for invalid instruction execution, we further check where the firmware loads the instruction address.
In the case of invalid instruction execution, the \instruction{PC} register, which holds the address of the next instruction, could be explicitly or implicitly modified.
For instructions that explicitly modify \instruction{PC} (e.g., \instruction{POP}),
the stack address from which \instruction{PC} loads the target address is identified as the taint sink.
For other instructions that implicitly modify \instruction{PC} (e.g., \instruction{BLX}), the used register is identified as the taint sink.

\begin{algorithm}[t]
    \SetAlgoLined
    \caption{Taint Propagation Algorithm}
    \label{algo:propagation-rule}
    \KwIn{$C$, Use-define chain that consists of use and define nodes}
    \KwIn{$t$, A register or memory address that identified as a taint sink}
    \KwOut{$T$, Tainted instruction set}
    $T=\phi$ \;
    $u\gets$ \textsc{MakeUseNode}($t$)\;
    $S=\{u\}$\;
    \While{\textsc{IsNotEmpty}($S$)}{
        $h\gets$ \textsc{GetAndRemoveElement}($S$)\;
        $T\gets T\cup$ \textsc{FindInstruction}($h$)\;
        $d\gets$ \textsc{FindDefineNode}($C, h$)\;
        $T\gets T\cup$ \textsc{FindInstruction}($d$)\;
        $U\gets$ \textsc{FindUseNodes}($C, d$)\;
        $S\gets S\cup U$\;
    }
\end{algorithm}

\noindent\textbf{Propagation rules.}
\system{} follows a systematic algorithm to propagate the taint backward through the use-define chain. 
Starting with the taint sink, it iteratively locates the define nodes and their corresponding use nodes, adding the associated instructions to the tainted set. 
This process continues until there are no more nodes to process.
Algorithm~\ref{algo:propagation-rule} shows how \system{} propagates the taint.
First, \system{} converts the taint sink to a use node and adds the node into an empty set $S$.
Then, for every node $u$ in $S$, \system{} leverages the use-define chain to locate its define node $d$, then removes $u$ from $S$.
The instructions associated with $d$ and $u$ are then tainted.
The data source of $d$ is used to locate its use nodes set $U$ in the use-define chain.
\system{} adds nodes set $U$ to $S$.
This backward tainting is executed repeatedly until there is no node in $S$.
By the end of the backward taint analysis, \system{} has identified the initial set of tainted instructions in $T$, which represent the potential root causes of the observed crash.

\subsubsection{Suspicious Score Assignment}\label{sec:rank-method}

Now we rethink the idea of automated root cause analysis.
The previous approach of utilizing a history-driven method to resolve memory aliases can result in a lengthy use-define chain.
Consequently, a large set of instructions may be reported as the potential root cause, which would place a significant burden on analysts to investigate manually.

To enhance the effectiveness of the postmortem-based fault localization method, we propose a heuristic approach to rank tainted instructions. 
Our insight lies in recognizing that instructions in the initial instruction set do not inherently possess equal importance. 
For instance, in the case of the \instruction{memcpy} function, which iterates through byte-by-byte copying, the repetition of identical instructions occurs frequently.
However, the instructions involving loading the target address and copying length hold greater significance, as they pertain to the initialization and termination conditions of the loop. 
Intuitively, such instructions should have higher ranks.
This aids analysts in prioritizing attention toward more critical instructions. 

\system{} assigns suspicious scores in two steps.
First, we assign an initial score to every tainted instruction.
Second, we heuristically model the behavior of assembly instructions and adjust their scores accordingly.
We primarily design a redundant loop taint suppression strategy from the perspective of control flow and a history write taint prioritization strategy from the perspective of data flow.
In the following, we detail the design of these two strategies.

\noindent\textbf{Redundant loop taint suppression.}
We consider consecutive instructions that are repeatedly executed as a redundant loop.
Take the aforementioned \instruction{memcpy} function as an example. 
The \instruction{memcpy} function requires three arguments, including two address pointers $ptr_1$ and $ptr_2$, and an integer $n$.
While executing, \instruction{memcpy} repeatedly copies one byte from $ptr_2$ to $ptr_1$ for $n$ times.
If $n$ is larger than the capacity of the data structure that $ptr_1$ or $ptr_2$ holds,
\instruction{memcpy} will read or write an illegal address, overwrite other variables, and trigger a firmware crash.
In this case, $n$ is usually a large number, resulting in an extremely long execution trace.
This leads to a lot of tainted instructions in backward tainting analysis.
Although these repetitive instructions are related to the crash, they are simply a sequence of value calculations and data copies that do not provide analysts with any insight into the crash.
Our insight is that, the repeatedly executed consecutive instructions are less important than the initialization and termination conditions of the loop.
Therefore, we decrease the suspicious score of repeatedly executed consecutive instructions.

\noindent\textbf{History write taint prioritization.}
Another critical behavior we consider is the history write, which refers to instructions that write to the exact memory address related to the crash site. 
Our observation is that for a deep root cause, the taint propagation can be extremely long, which taints many instructions. 
In such cases, the taint source instructions,which perform history writes and are far away from the taint sink, are more important than others along the taint propagation path. 
Therefore, we increase the suspicious scores of these taint source instructions.

Finally, \system{} sorts all tainted instructions according to their suspicious scores. 
Instructions with higher scores are given greater priorities and ranks, allowing analysts to focus their attention on the most relevant instructions during the root cause investigation.
\section{Implementation}

We implement the prototype of \system{} for the purpose of automatically analyzing the root cause of firmware crashes on the 32-bit ARM Cortex-M architecture.
\system{} mainly consists of three components, including footprint collection, reverse execution, and root cause analysis.
In the following, we introduce the implementation details of these components.

\noindent\textbf{Footprint collection.}
To capture the desired runtime data during emulation, we develop two kinds of events on top of Unicorn's~\cite{unicorn} hooking mechanism. 
Specifically, we implement conditional callbacks for these two events, including \hookMR{} and \hookMW{} for data events, and \hookE{} for action events.
By registering these callbacks, we are able to precisely monitor and record the relevant program behaviors and data accesses during the crash reproducing process.

\noindent\textbf{Reverse execution.}
To analyze the raw binary firmware, we develop an instruction analyzer on top of the widely adopted Capstone disassembly framework \cite{capstone}. 
Capstone provides semantic details about the disassembled instructions, including implicit register reads and writes. 
We design a handler to examine the source and destination operands of each instruction in action events, then create corresponding use and define nodes in the program representation. 
When the instruction accesses memory locations that have been previously captured by our data event monitoring, our handler is able to extract the concrete address and data values. 
This eliminates the need for expensive memory alias resolution, allowing accurate analysis of the full execution trace.
Besides, we implement \TotalHandlers{} unique instruction handlers to perform inverse transformations on registers to facilitate reverse execution.

\noindent\textbf{Root cause analysis.}
To identify the critical program points that could lead to potential crashes, we develop a set of taint sink rules covering all relevant instructions. 
By leveraging our ability to precisely track memory accesses, as described earlier, we are able to apply these taint rules effectively. 
However, this also induces the risk of over-tainting, where excessive numbers of instructions get flagged as potentially suspicious.
To address this challenge, we design a novel heuristic ranking strategy to highlight the most instructive taint sinks. 
The ranking strategy, as described in Section~\ref{sec:rank-method}, allows \system{} to focus on reporting the instructions that provide the most valuable guidance for understanding the observed crashes. 

\section{Evaluation}
In this section, we assess the performance and capabilities of our \system{} framework from multiple perspectives. 
First, we examine whether \system{} can accurately identify the root causes of observed firmware crashes. 
Next, we measure the time breakdown of each component of \system{} and the overall efficiency. 
Then, we conduct an ablation study to measure the contribution of our heuristic ranking strategy. 
Finally, we evaluate the event-based design to demonstrate its contribution to \system{}.

We compare \system{} against two latest postmortem-based root cause analysis works, POMP \cite{xu2017postmortem} and POMP++ \cite{mu2019pomp++}.
We choose these works because they and \system{} consider the semantics of instructions, which provide more instructive results for analysts than spectrum-based methods\cite{sarhan2022survey}.
POMP performs backward taint analysis on a core dump file and the execution trace collected by Intel PT\cite{intelpt} to highlight the root cause of a crash.
POMP++ further enhances POMP in memory alias resolving with value-set analysis.
While these approaches are designed for the x86 Linux platform, we have implemented them on the 32-bit ARM Cortex-M architecture to ensure a fair comparison.
We detail our implementation for these works and the measures taken to ensure a fair comparison with them in Appendix~\ref{app:migration}.
Our evaluation demonstrates the significant advancements that \system{} brings over these state-of-the-art works in terms of overall accuracy, efficiency, and full execution trace analysis capability.
In summary, we evaluate \system{} with the following research questions:

\noindent \textbf{RQ1:} What is the ability of \system{} to identify the root cause of an embedded firmware crash? (Section \ref{sec:effectiveness})

\noindent \textbf{RQ2:} What is the overhead of every component of \system{}, and how is the overall efficiency? (Section \ref{sec:efficiency}) 

\noindent \textbf{RQ3:} What is the contribution of heuristic ranking strategies? (Section \ref{sec:ablation})

\noindent \textbf{RQ4:} How does the event-based design contribute to \system{}'s performance? (Section \ref{sec:component-compare})

\begin{table}[ht]
\centering
\caption{Comparison result of POMP (\pomp{}), POMP++ (\pompplus{}) and \system{} (F) on analyzing full instructions (Full) and the last 50\% instructions (Half), where $\varnothing$ indicates the analysis timed out, \xmark{} indicates the root cause is not found, and a number represents the rank of the root cause instruction in the result. Failures in the \system{} analysis of full instructions are \underline{underlined}. $\Delta\mathrm{Root}$ denotes the distance between the root cause and crash site, followed by a proportion in full execution trace. ID is highlighted in \colorbox{gray!20}{GRAY} if its $\Delta\mathrm{Root}$ is larger than 50\%. \# Traces denotes the total number of instructions in the complete execution trace. \# Ins denotes the total number of instructions in the firmware. }
\resizebox{1\linewidth}{!}{\begin{tabular}{@{}crrrccccccc@{}}\toprule
\multicolumn{1}{c}{\multirow{2}{*}{ID}} &
  \multicolumn{1}{c}{\multirow{2}{*}{$\Delta\mathrm{Root(\%)}$}} &
  \multicolumn{1}{c}{\multirow{2}{*}{\# Traces}} &
  \multicolumn{1}{c}{\multirow{2}{*}{\# Ins}} &
  \multicolumn{3}{c}{Full} &
   &
  \multicolumn{3}{c}{Half} \\ \cmidrule(lr){5-7} \cmidrule(l){9-11} 
\multicolumn{1}{c}{} &
  \multicolumn{1}{c}{} &
  \multicolumn{1}{c}{} &
  \multicolumn{1}{c}{} &
  \multicolumn{1}{c}{\pomp{}} &
  \multicolumn{1}{c}{\pompplus{}} &
  \multicolumn{1}{c}{F} &
   &
  \multicolumn{1}{c}{\pomp{}} &
  \multicolumn{1}{c}{\pompplus{}} &
  \multicolumn{1}{c}{F} \\ \midrule 
$C_{1}$ &  2 (<0.1\%) &52,080 &11,869 & $\varnothing$ & \xmark{} & 1 & & \xmark{} & \xmark{} & 1 \\ 
$C_{2}$ &  2 (<0.1\%) &371,762 &13,374 & $\varnothing$ & $\varnothing$ & 1 & & $\varnothing$ & $\varnothing$ & 1 \\ 
$C_{3}$ &  2 (<0.1\%) &1,003,421 &13,374 & $\varnothing$ & $\varnothing$ & 1 & & $\varnothing$ & $\varnothing$ & 1 \\ 
$C_{4}$ &  2 (<0.1\%) &228,012 &5,021 & $\varnothing$ & \xmark{} & 1 & & $\varnothing$ & \xmark{} & 1 \\ 
$C_{5}$ &  2 (<0.1\%) &36,271 &11,880 & \xmark{} & \xmark{} & 1 & & \xmark{} & \xmark{} & 1 \\ 
$C_{6}$ &  2 (<0.1\%) &108,294 &7,126 & $\varnothing$ & \xmark{} & 1 & & $\varnothing$ & \xmark{} & 1 \\ 
$C_{7}$ &  2 (<0.1\%) &619,937 &15,140 & $\varnothing$ & $\varnothing$ & 1 & & $\varnothing$ & $\varnothing$ & 1 \\ 
$C_{8}$ &  2 (<0.1\%) &68,894 &15,513 & $\varnothing$ & \xmark{} & 1 & & \xmark{} & \xmark{} & 1 \\ 
$C_{9}$ &  2 (<0.1\%) &524,897 &11,931 & $\varnothing$ & $\varnothing$ & 1 & & $\varnothing$ & $\varnothing$ & 1 \\ 
$C_{10}$ &  20,203 (18.1\%) &111,898 &18,536 & $\varnothing$ & \xmark{} & 2 & & \xmark{} & \xmark{} & 2 \\ 
$C_{11}$ &  78 (<0.1\%) &168,542 &17,165 & $\varnothing$ & 1 & 1 & & 1 & 1 & 1 \\ 
\cellcolor{gray!20}$C_{12}$ &  120,472 (51.1\%) &235,662 &9,299 & $\varnothing$ & \xmark{} & \underline{13} & & \xmark{} & \xmark{} & 8 \\ 
$C_{13}$ &  2,176 (1.4\%) &160,721 &9,299 & $\varnothing$ & $\varnothing$ & \underline{21} & & $\varnothing$ & \xmark{} & 19 \\ 
$C_{14}$ &  8,155 (39.5\%) &20,630 &15,428 & \xmark{} & \xmark{} & 1 & & \xmark{} & \xmark{} & 1 \\ 
\cellcolor{gray!20}$C_{15}$ &  8,719 (73.0\%) &11,943 &17,165 & \xmark{} & \xmark{} & 1 & & \xmark{} & \xmark{} & \xmark{} \\ 
\cellcolor{gray!20}$C_{16}$ &  215,102 (98.7\%) &217,900 &17,165 & $\varnothing$ & \xmark{} & 10 & & $\varnothing$ & \xmark{} & \xmark{} \\ 
$C_{17}$ &  5,183 (3.0\%) &173,913 &17,165 & $\varnothing$ & 8 & 1 & & \xmark{} & 8 & 1 \\ 
$C_{18}$ &  2,759 (1.6\%) &169,666 &17,165 & $\varnothing$ & \xmark{} & 1 & & \xmark{} & \xmark{} & 1 \\ 
\cellcolor{gray!20}$C_{19}$ &  5,161 (72.2\%) &7,151 &9,299 & \xmark{} & \xmark{} & 1 & & \xmark{} & \xmark{} & \xmark{} \\ 
$C_{20}$ &  306 (<0.1\%) &523,873 &14,325 & $\varnothing$ & \xmark{} & 1 & & \xmark{} & \xmark{} & 1 \\ 
$C_{21}$ &  2,013 (1.3\%) &150,995 &19,706 & $\varnothing$ & \xmark{} & 2 & & \xmark{} & \xmark{} & 2 \\ 
$C_{22}$ &  401 (0.2\%) &238,470 &13,954 & $\varnothing$ & \xmark{} & 2 & & $\varnothing$ & \xmark{} & 2 \\ 
$C_{23}$ &  53 (<0.1\%) &89,136 &32,546 & $\varnothing$ & \xmark{} & 1 & & $\varnothing$ & \xmark{} & 1 \\ 
$C_{24}$ &  53 (0.1\%) &52,909 &32,546 & $\varnothing$ & \xmark{} & 1 & & \xmark{} & \xmark{} & 1 \\ 
$C_{25}$ &  1,789 (0.4\%) &434,172 &13,954 & $\varnothing$ & 1 & 2 & & $\varnothing$ & 1 & 2 \\ 
$C_{26}$ &  88,746 (45.2\%) &196,313 &29,993 & $\varnothing$ & 3 & 1 & & 3 & 3 & 1 \\ 
\cellcolor{gray!20}$C_{27}$ &  27,411 (58.2\%) &47,065 &15,529 & 3 & 3 & 3 & & 3 & 3 & 4 \\ 
\cellcolor{gray!20}$C_{28}$ &  28,771 (58.9\%) &48,880 &15,529 & 2 & 2 & 2 & & 2 & 2 & 3 \\ 
\cellcolor{gray!20}$C_{29}$ &  55,768 (66.4\%) &84,028 &23,706 & $\varnothing$ & \xmark{} & 1 & & \xmark{} & \xmark{} & \xmark{} \\ 
\cellcolor{gray!20}$C_{30}$ &  59,717 (63.3\%) &94,362 &23,706 & $\varnothing$ & \xmark{} & 1 & & \xmark{} & \xmark{} & \xmark{} \\ 
$C_{31}$ &  402 (<0.1\%) &514,772 &14,325 & \xmark{} & \xmark{} & 1 & & \xmark{} & \xmark{} & 1 \\ 
$C_{32}$ &  86,956 (8.7\%) &994,960 &27,664 & $\varnothing$ & $\varnothing$ & \underline{\xmark{}} & & $\varnothing$ & $\varnothing$ & \xmark{} \\ 
$C_{33}$ &  100,512 (48.2\%) &208,669 &18,464 & $\varnothing$ & 2 & 2 & & $\varnothing$ & 2 & 3 \\ 
$C_{34}$ &  21 (<0.1\%) &160,958 &18,475 & $\varnothing$ & \xmark{} & 1 & & $\varnothing$ & \xmark{} & 1 \\ 
$C_{35}$ &  337 (<0.1\%) &1,960,419 &26,165 & $\varnothing$ & $\varnothing$ & 1 & & $\varnothing$ & $\varnothing$ & 1 \\ 
$C_{36}$ &  274,858 (27.1\%) &1,016,078 &26,167 & $\varnothing$ & $\varnothing$ & 1 & & $\varnothing$ & $\varnothing$ & \xmark{} \\ 
\cellcolor{gray!20}$C_{37}$ &  1,458,974 (99.0\%) &1,474,404 &18,202 & $\varnothing$ & $\varnothing$ & 1 & & $\varnothing$ & $\varnothing$ & \xmark{} \\ 
$C_{38}$ &  44,256 (3.4\%) &1,300,470 &25,887 & $\varnothing$ & $\varnothing$ & 1 & & $\varnothing$ & $\varnothing$ & 1 \\ 
\cellcolor{gray!20}$C_{39}$ &  42,177 (53.6\%) &78,624 &15,605 & $\varnothing$ & \xmark{} & 1 & & \xmark{} & \xmark{} & 1 \\ 
$C_{40}$ &  891 (1.7\%) &52,173 &11,952 & 4 & 4 & 5 & & 4 & 4 & 5 \\ 
\cellcolor{gray!20}$C_{41}$ &  594,471 (95.1\%) &625,021 &12,258 & $\varnothing$ & $\varnothing$ & 1 & & $\varnothing$ & \xmark{} & \xmark{} \\ 
\bottomrule
\end{tabular}}
\label{tbl:summary}
\end{table}

\vspace{-8pt}
\subsection{Experimental Setup}
\vspace{-4pt}

\noindent\textbf{Experiment settings.}
We perform the evaluation under the same experiment setting: a 56-core Intel(R) Xeon(R) CPU E5-2680 v4 @ 2.40GHz machine running Ubuntu 22.04.2 LTS with 256GB RAM. 

\noindent\textbf{Dataset and ground truth.}
To conduct a comprehensive evaluation, we utilize the public dataset provided by the Fuzzware framework \cite{fuzzware-dataset}. 
This dataset contains 61 test suites that cover a wide range of crashing behaviors.
Each test suite includes the raw binary firmware, a basic configuration file, a crashing test case, and a crash description. 
We follow the official guidance of Fuzzware to deploy the framework in its reproducing mode and successfully reproduce \TotalTestCases{} test cases that exhibit crashing behaviors. 
We assign crash IDs from $C_{1}$ to $C_{41}$ for these test cases, which are detailed in Table~\ref{tbl:evaluation-dataset} in Appendix~\ref{app:crash-id}.
For each of these test suites, we employ a two-pronged approach to establish the ground truth. 
First, we extract the instructions identified in the provided crash descriptions as the root causes. 
Then, for those test suites lacking explicit descriptions, we manually inspect the assembly code of their firmware images and leverage execution logs to identify root causes.

\subsection{Effectiveness of \system{} (RQ1)}\label{sec:effectiveness}

\noindent\textbf{Root cause analysis ability.} 
We provide each analyzer with the crashing test case and the corresponding firmware to analyze and set a timeout of 48 hours.
If the faulty instruction is reported in the \topten{} positions in the fault localization result, we consider it a success \cite{kochhar2016practitioners}.
Table~\ref{tbl:summary} shows the root cause analysis ability of \system{} (F) compared to POMP (\pomp{}) and POMP++ (\pompplus{}).
$\Delta\mathrm{Root}$ represents the number of instructions executed between the root cause and the crash site, along with its proportion in the complete execution trace of the crashing test case. 
We first use the full execution trace of the crashing test case to analyze the root cause.
The results show that \system{} is able to successfully identify the root cause in \SystemSuccessNum{} out of \TotalTestCases{} test cases, achieving a success rate of \SystemSuccessRatio{}. 
We underline the failed results (i.e., $C_{12}$, $C_{13}$ and $C_{32}$) in the full execution trace setting.
POMP and POMP++ only identify 3 and 8 out of all test cases, with success rates of 7.3\% and 19.5\%, respectively.
POMP fails in the majority of test cases due to the huge time demands that exceed the timeout.
POMP++ fails because of the inadequacy of its heuristic approaches to resolving memory aliasing, which proves unsuitable for the analysis of embedded firmware.
In the full execution trace setting, \system{} is able to complete analysis on all test cases without any timeouts.
In contrast, POMP and POMP++ only finish 19.5\% and 73.2\% of test cases, respectively.
To enable a fair comparison, the execution trace under analysis is then limited to the last 50\% instructions before the crash site. 
In this limited setting, \system{} still successfully identifies 31 test cases with a success rate of 75.6\%, outperforming POMP (12.2\% success rate) and POMP++ (19.5\% success rate).

\noindent\textbf{\system{} failure analysis.}
While \system{} is able to successfully identify the root causes in most test cases, it fails on \SystemFailedCases{} test cases.
\system{} fails to identify the root causes of $C_{12}$ and $C_{13}$ within \topten{} instructions because their execution behaviors related to the crashes do not satisfy the heuristic suspicious score assignment strategy of \system{}.
We further discuss the effectiveness of our strategy in the ablation study in Section~\ref{sec:ablation}.
The main reason for the failure of $C_{32}$ is that the underlying assumption of \system{}'s approach is not met.
In this case, an already corrupted data section leads to a crashed interrupt context restore during an interrupt context switch process.
Our event-based footprint collection method fails to collect the execution trace of the interrupt context restore process.
Thus, \system{} fails to identify the taint sink and the root cause.
However, this is an isolated incident and does not impact the overall performance of \system{}.
We discuss possible mitigations in Section~\ref{sec:future-work}.

\noindent\textbf{Deep root cause analysis.}
The root cause of a crash may be far away from the crash site, which we call it a deep root cause.
For example, there are 98.7\% of all instructions executed before crash in $C_{16}$.
A deeper root cause introduces a longer execution trace between the root cause and the crash site, resulting in more unresolved memory aliases that complicate the analysis.
Unresolved memory aliases often lead to considerable analysis time costs and can even result in incorrect fixes, posing challenges for both automated analysis tools and manual investigation.
Table~\ref{tbl:summary} shows the effectiveness of deep root cause analysis.
Taking the $\mathrm{\Delta Root}$ over \DeepRCABar{} as deep root causes, whose crash IDs are marked in gray, \system{} succeeds in \DeepRCASucess{} out of \DeepRCATotal{} test cases within \topten{} instructions.
Take \DeepRCASample{} for example, a buffer overflow in the \instruction{memcpy} function results in memory corruption.
The third argument of \instruction{memcpy}, which specifies the number of bytes to copy, is set to a large value, leading to an extremely long execution trace and a deep root cause.
\system{} is able to successfully identify the root cause of the \DeepRCASample{} crash, with a root cause analysis depth of \DeepRCADepth{} instructions.

\begin{figure*}[tb]
    \centering
    \setlength{\belowcaptionskip}{-2mm}
    \includegraphics[width=1\textwidth]{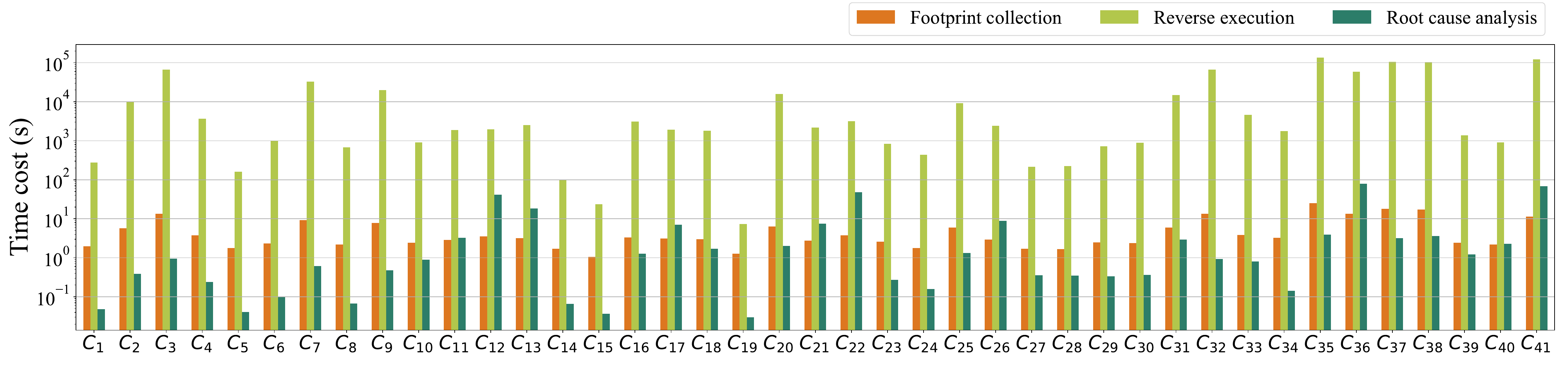} 
    \caption{Time overhead breakdown of \system{} in a log scale. }
    \label{fig:time-breakdown} 
\end{figure*}

\begin{takeaway}
    \textbf{RQ1:}  
    \system{} identifies the root cause of \SystemSuccessNum{} out of \TotalTestCases{} test cases with a successful rate of \SystemSuccessRatio{}, outperforming POMP (7.3\% success rate) and POMP++ (19.5\% success rate), demonstrating its capability to analyze the full execution trace. %in full execution trace setting.
    \system{} is able to localize deep root causes and succeeds in 10 out of 11 crashing test cases whose root causes are located more than 50\% away from the crash site along the full execution trace.
\end{takeaway}

\subsection{Efficiency of \system{} (RQ2)}\label{sec:efficiency}

In this section, we first measure the overhead of every component of \system{}.
Then, we measure the overall efficiency.

\noindent\textbf{Footprint collection overhead.}
To measure the overhead of event-based footprint collection approach, we repeatedly reproduce each crashing test case both with and without the added events 50 times. 
The results are shown in Table~\ref{tbl:events-overhead-summary}, and the details are shown in Table~\ref{tbl:events-overhead} in Appendix~\ref{app:hook-overhead}. 
The event-based footprint collection method introduces an average time of \AvgOverheadTimes{} and an average log file size of \AvgOverheadSpace{}.
It is worth noting that \system{} performs the event-based footprint collection as an independent step after the fuzzing process.
This means that the overhead associated with the event-based approach does not impact the overall efficiency of the fuzzing process itself.

%%%%%%%%%%%%%%%%%%%%%%%%%%%%%%%%%%%%%%%%%%%%%%%%%%%%
%Both

\begin{table}[ht]
    \centering
    \caption{Statistical results of footprint collection overhead.}
    \resizebox{1\linewidth}{!}{
        \begin{tabular}{@{}clcccc@{}}
        \toprule
        \multicolumn{1}{l}{}      &            & No Events & Data Events & Action Events & Both Events \\ \midrule
        \multirow{2}{*}{Time (s)} & Average    & 0.82      & 2.63        & 4.07          & 5.47        \\ \cmidrule(l){2-6} 
                          & Addition &  -       & 2.21$\times{}$   & 3.96$\times{}$   & 5.67$\times{}$ \\ \midrule
File Size (MB)                & Average    & 0         & 9.51        & 13.62         & 23.16       \\ \bottomrule
        \end{tabular}
    }
    \label{tbl:events-overhead-summary}
\end{table}

\noindent\textbf{Time overhead breakdown.}
We measure the execution time of every component of \system{} to assess its impact on the overall overhead.
For each crashing test case, \system{} performs fault localization in the full execution trace setting with a timeout of 48 hours.
Figure~\ref{fig:time-breakdown} illustrates the result.
First, while the event-based footprint collection incurs some additional overhead, it provides the runtime information that is crucial for \system{}'s comprehensive root cause analysis. 
This finding suggests that the trade-off between the overhead and the benefits of the event-based footprint collection is actually more favorable than previous works assumed \cite{cui2018rept,xu2017postmortem,xuracing}. 
Second, the reverse execution occupies the majority of the time cost, indicating that it is worth optimizing.
We further demonstrate the contribution of our event-based data recovery to reverse execution in Section~\ref{sec:component-compare}.
Finally, the time overhead of the root cause analysis component varies mainly due to the number of tainted instructions.
\system{}'s backward taint analysis and suspicious score assignment introduce only a little overhead with an average of 7.66s.

\begin{figure}[htbp]
    \centering
    \setlength{\belowcaptionskip}{-2mm}
    \includegraphics[width=0.45\textwidth]{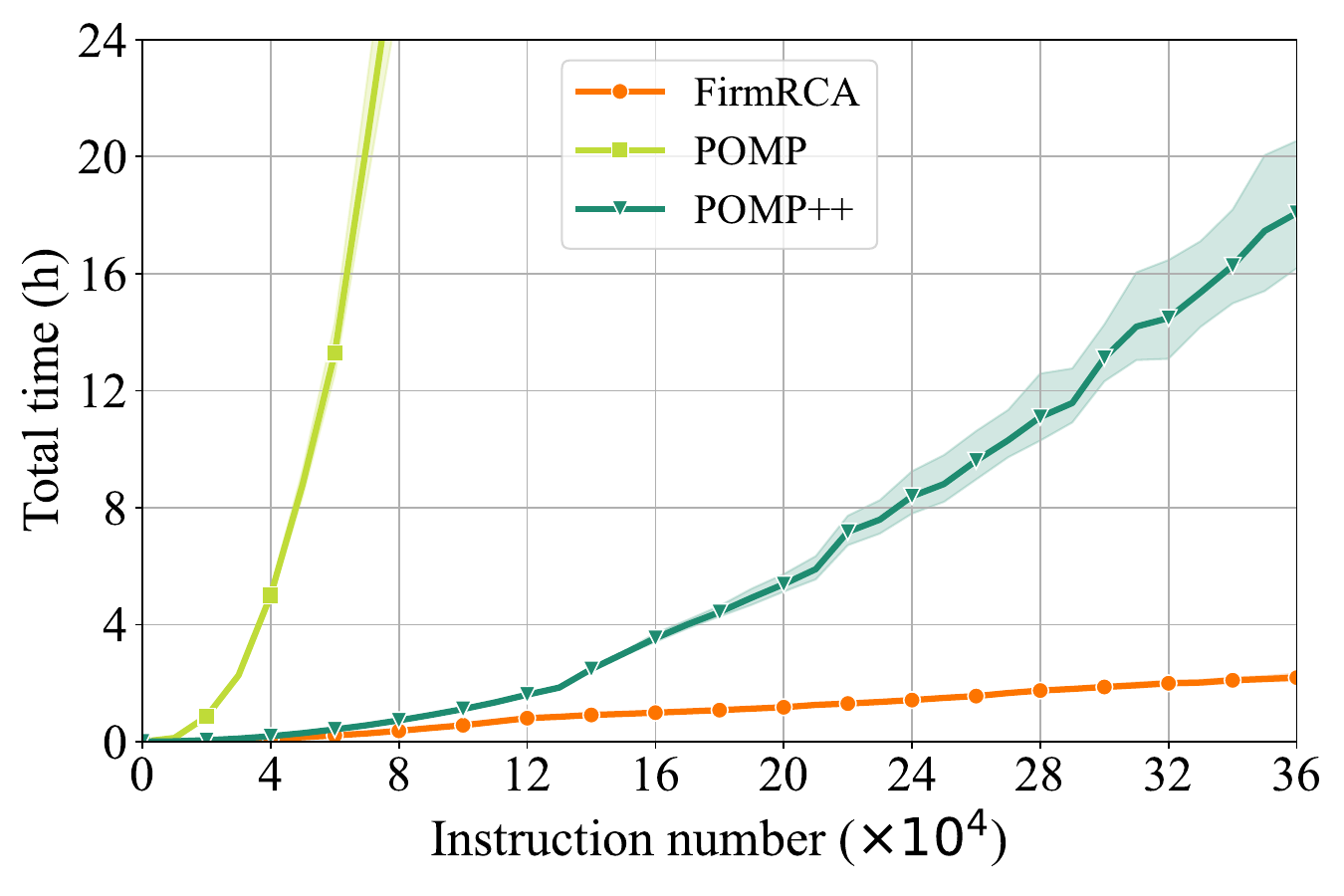} 
    \caption{Overall time costs of fault localization on the different number of instructions under analysis. }
    \label{fig:time-growth} 
\end{figure}

\begin{figure*}[htbp]
    \centering
    \setlength{\belowcaptionskip}{-1mm}
    \includegraphics[width=1\textwidth]{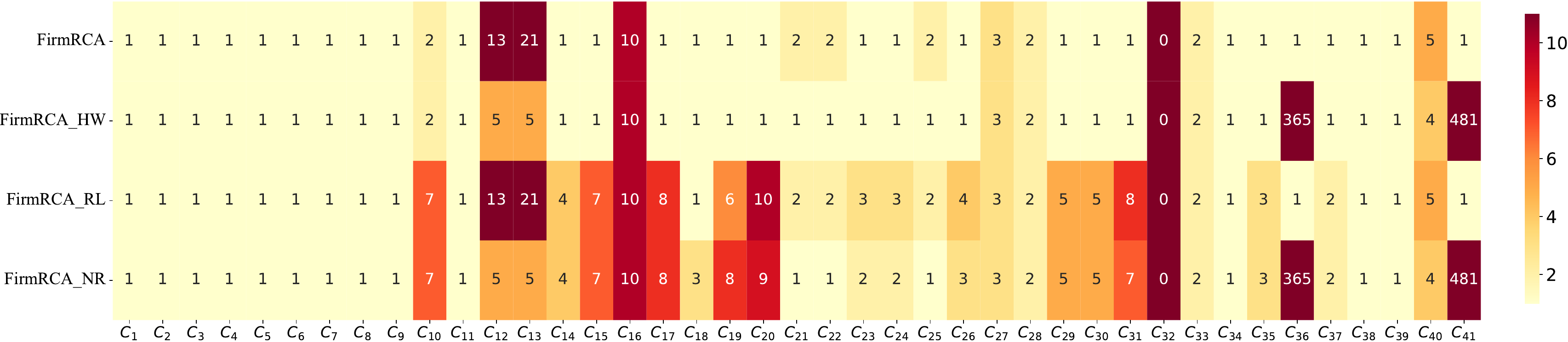} 
    \caption{Comparison results of root cause identification of \SystemNoRanking{}, \SystemAblationRL{}, \SystemAblationHW{} and \system{} in the heatmap visualization, where lighter shades indicate better results. The number in the cell represents the rank of the root cause instruction in the result. }
    \label{fig:ranking-strategy} 
\end{figure*}

\noindent\textbf{Overall time costs.}
As discussed in Section~\ref{sec:effectiveness}, the length of the execution trace under analysis has a significant impact on the efficiency of the root cause analysis.
In this work, we define the analysis depth as the number of instructions analyzed leading up to the crash site.
A larger value of analysis depth can lead to more unresolved memory aliases, resulting in an increase in the number of use nodes and define nodes whose values are unknown in the constructed use-define chain.
These unresolved nodes not only induce extra overhead when updating the use-define chain, but they also limit the ability to perform deep root cause analysis. 
To demonstrate this impact, we randomly select 5 crashing test cases with more than 360,000 instructions in execution traces and calculate average overall time costs with a timeout of 24 hours.
Figure~\ref{fig:time-growth} shows the result of one of the test cases ($C_{38}$). The result of the other four test cases is shown in Figure~\ref{fig:multiple-time-growth} in Appendix~\ref{app:multiple-time-growth}. 
As the analysis depth increases linearly, POMP exhibits an exponential growth trend, while POMP++ exhibits a polynomial-level growth trend.
In contrast, \system{} demonstrates only a modest overall time cost as the analysis depth increases.
This difference in performance highlights a key advantage of \system{}'s event-based footprint collection that helps resolve memory aliases. 
\system{} minimizes the impact of unresolved memory aliases, and thus is able to perform a much more efficient fault localization, even for cases with extremely long execution traces.
This capability is particularly valuable in complex firmware crash scenarios, where the root cause can be deeply hidden in the execution trace.

\begin{takeaway}
    \textbf{RQ2:}  
    \system{}'s event-based footprint collection approach introduces an average additional time of \AvgOverheadTimes{} and an average additional log file size of \AvgOverheadSpace{}.
    \system{} demonstrates its significant improvement in the overall efficiency of fault localization, outperforming POMP and POMP++.

\end{takeaway}

%%%%%%%%%%%%%%%%%%%%%%%%%%%%%%%%%%%%%%%%%%%%%%%%%%%%%%%
%
%  RQ 3
%
%%%%%%%%%%%%%%%%%%%%%%%%%%%%%%%%%%%%%%%%%%%%%%%%%%%%%%%

\subsection{Efficacy of Ranking Strategies (RQ3)}\label{sec:ablation}

To quantify the impact of our heuristic ranking strategies, we construct four prototypes of the \system{} framework:
(1) \SystemNoRanking{} as the basic version that has no ranking strategies enabled;
(2) \SystemAblationRL{} that has only the redundant loop taint suppression strategy enabled;
(3) \SystemAblationHW{} that has only the history write taint prioritization strategy enabled;
(4) \system{} that incorporates both ranking strategies.
For each of the \TotalTestCases{} crashing test cases in our evaluation dataset, we set a 48-hour analysis timeout in the full execution trace setting. 
As shown in Figure~\ref{fig:ranking-strategy}, all the prototypes successfully identify the root causes of \SystemSuccessNum{} out of \TotalTestCases{} test cases within \topten{} ranked instructions, achieving a success rate of \SystemSuccessRatio{}. 

\noindent\textbf{Redundant loop taint suppression strategy.}
Compared to \SystemNoRanking{}, \SystemAblationRL{} highlights the root causes of the $C_{36}$ and $C_{41}$ test cases at \topone{} instruction.
Take $C_{41}$ for example, the \SystemNoRanking{} taints hundreds of \instruction{ADDS	R0, \#0x40} instructions along the execution trace.
These simple value calculations continuously add 0x40 to \instruction{R0}, which are useless for analysts to investigate the root cause of the crash.
The redundant loop taint suppression strategy then decreases their suspicious scores and successfully identifies the root cause.

\noindent\textbf{History write taint prioritization strategy.}
Compared to \SystemNoRanking{}, \SystemAblationHW{} further improves ranking results in 15 (36.6\%) test cases ($C_{10}$, $C_{14-15}$, $C_{17-20}$, $C_{23-24}$, $C_{26}$, $C_{29-31}$, $C_{35}$, $C_{37}$). 
Besides, \SystemAblationHW{} achieves the best performance in identifying the root causes at \topone{} instruction of 73.2\%, outperforming \SystemAblationRL{} (39\%), \SystemNoRanking{} (39\%) and \system{} (70.7\%).

\noindent\textbf{Overall performance.}
\system{} adopts a dual strategy approach to achieve optimal performance.
On one hand, \system{} benefits from the redundant loop taint suppression strategy in $C_{36}$ and $C_{41}$.
On the other hand, \system{} retains the improvement on the aforementioned 15 test cases with the help of the history write taint prioritization strategy.
However, we also observe some side effects in $C_{12}$ and $C_{13}$ resulting from \SystemAblationRL{}.
For these two test cases, their root causes are contained in a loop of instructions before crashes, leading to a downside in final ranking results.
\update{Compared with \SystemNoRanking{}, the average of the degradation in \SystemAblationRL{} is only 3 instructions.
In the final manual investigation, we consider this slight performance degradation to be acceptable.}

\begin{takeaway}
    \textbf{RQ3:}  
    All four prototypes demonstrate a successful identification of the root causes in \SystemSuccessNum{} out of \TotalTestCases{} test cases within \topten{} ranked instructions, achieving a success rate of \SystemSuccessRatio{}.
    The redundant loop taint suppression strategy helps level up the rank to \topone{} and the history write taint prioritization strategy enhances the ranking results in 36.6\% of the test cases.
\end{takeaway}

\vspace{-4pt}
\subsection{Event-based Method Contribution (RQ4)}\label{sec:component-compare}
\vspace{-4pt}

\begin{figure*}[tb]
    \centering
    \includegraphics[width=1\textwidth]{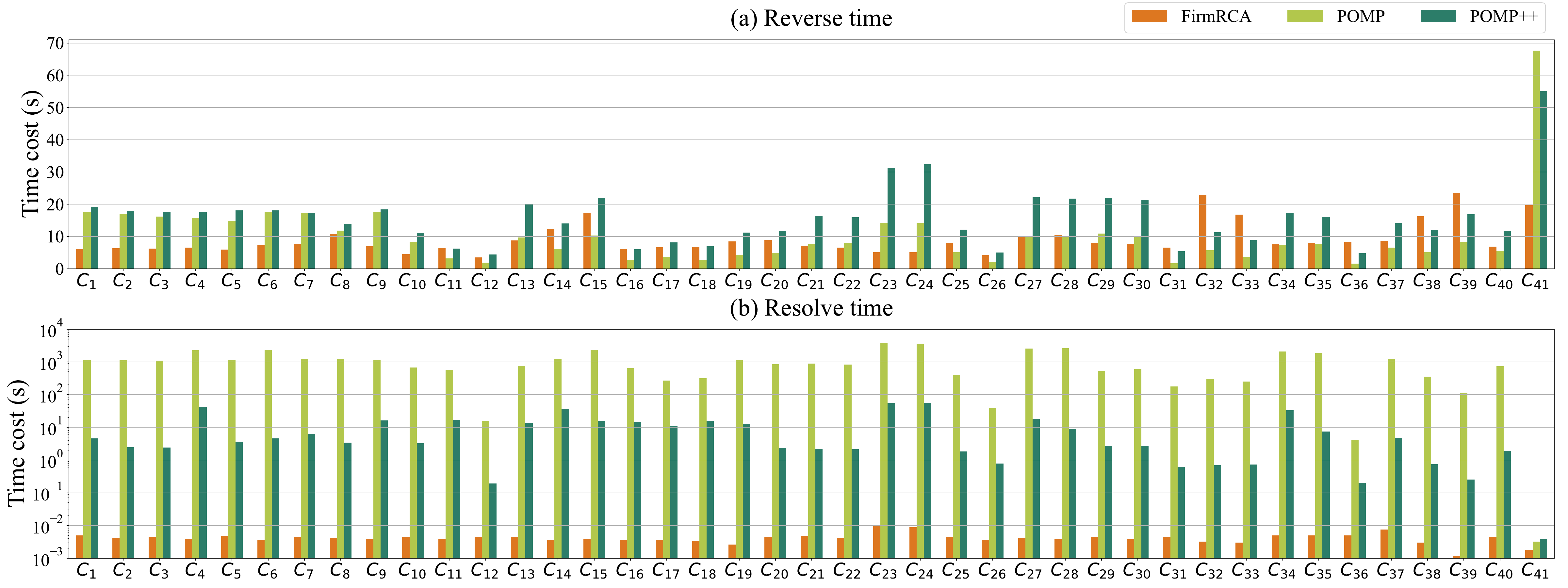} 
    \caption{The time overhead breakdown of the reverse execution on \TimeBreakDownDepth{} among \system{}, POMP and POMP++. (a) Reverse time shows use-define chain construction time cost. (b) Resolve time shows the data recovery time cost in a log scale.}
    \label{fig:component-compare} 
\end{figure*}

To further investigate the contribution of our event-based method of \system{}, we measure the time overhead breakdown of the reverse execution component in our design, including use-define chain construction and event-based data recovery.
For the sake of consistency, we split POMP and POMP++ into these two steps and rename the steps as \textit{reverse} and \textit{resolve}, respectively.
In the reverse period, each analyzer performs use-define chain construction along the execution trace.
In the resolve period, the analyzer inversely executes the instruction and tries to resolve memory aliases for data recovery.

We conduct this evaluation at an analysis depth of \TimeBreakDownDepth{}, where all experiments are able to finish within a reasonable time budget (i.e., 24 hours).
We repeatedly evaluate each test case for 5 times. 
Figure~\ref{fig:component-compare} illustrates the result of the experiment.
On average, POMP and POMP++ spend  \POMPReverseAvgTime{} and \POMPPlusReverseAvgTime{} in the reverse period and \POMPResolveAvgTime{} and \POMPPlusResolveAvgTime{} in resolve period, respectively.
In comparison, \system{} leverages its event-based data recovery approach to resolve memory aliases with significantly reduced time costs in the resolve period, averaging just \SystemResolveAvgTime{}. 

\begin{takeaway}
    \textbf{RQ4:}  
    \system{}'s event-based data recovery approach in the reverse execution phase greatly optimizes memory aliases resolve.
    This substantial improvement in the efficiency of memory alias resolve is a key contributor to \system{}'s overall enhanced performance in fault localization.
\end{takeaway}
\section{Discussion}

Based on the evaluation results, we provide the following insights into postmortem-based fault localization methods in embedded firmware fuzzing.

\noindent\textbf{Memory alias resolve problem.}
Recent postmortem-based works hold the view that recording runtime data introduces huge time overheads \cite{xu2017postmortem, guo2019deepvsa}.
Consequently, they infer memory aliases relying on core dumps, the execution trace, and a memory snapshot at the crash site.
The difference between inferred data and actual data hinders the performance of reverse execution and backward taint analysis.
In contrast, \system{} employs the event-based footprint collection to accurately resolve memory aliases and perform efficient fault localization.

\noindent\textbf{Event recording.}
In the fault localization for embedded firmware fuzzing scenario, recording the execution trace and memory access history greatly improves efficiency and effectiveness with little overhead.
For one thing, the event-based footprint collection method compensates for inadequate debugging mechanisms.
For another, we trigger data events and action events only when reproducing crashing test cases, which is an independent period after fuzzing and does not negatively impact the fuzzing process.

\noindent\textbf{Hardware tracing.}
On real hardware devices, analysts can also use a tracing unit to collect the runtime information.
For example, with the support of the Embedded Trace Macrocell (ETM\cite{etm}) feature, analysts are able to obtain the execution trace \cite{zhang2023alligator}.
However, these features require hardware settings to be enabled and do not exhibit priorities against our event-based footprint collection in terms of quantity and quality.
In this work, \system{} utilizes the event-based footprint collection also for the sake of scalability and financial cost.

\noindent\textbf{Execution trace length.}
Given a new crashing test case, analysts initially lack knowledge of the precise location where the root cause resides.
Intuitively, to accurately identify the root cause that could be far from the crash site, a longer trace would be preferable.
However, as the number of instructions to analyze grows, the cost of the analysis time increases rapidly.
When dealing with execution traces of tens of thousands of instructions, the total analysis period could even stretch to several days.
Therefore, it is important for postmortem-based fault localization methods to determine the appropriate length of the execution trace to analyze.
\system{} is able to complete analysis under the full execution trace setting in a reasonable time budget, demonstrating its efficiency and capability.

\section{Limitations and Future Work}
While \system{} has demonstrated great performance in identifying the root causes of firmware crashes, there are some limitations that have not been fully addressed.
In this section, we discuss the limitations of \system{} and possible avenues for future work.

\subsection{Limitations}\label{sec:limitations}

\noindent\textbf{Fuzzing constraints.}
\system{} aims to identify faulty instructions in the fuzzing scenario, serving as a post-fuzzing work.
The capabilities of the underlying fuzzer constrain the range of firmware and platforms because there are no perfect firmware fuzzers available to precisely emulate all kinds of firmware images.
This imposes certain limitations on the range of firmware that can be effectively tested. 

\noindent\textbf{Report usability.}
We target raw firmware binaries where all debugging symbols are removed for better scalability.
\system{} reports the ranked instructions to analysts, but does not mean to get rid of all the manual investigation in the root cause analysis work.
However, \system{} highlights key instructions related to the crash from tens of thousands of instructions in the execution trace, providing a crucial starting point for further manual analysis.

\noindent\textbf{Ranking strategy generalizability.}
\system{} employs two heuristic ranking strategies to perform suppression and prioritization on suspicious scores primarily based on behaviors of instructions. 
Although these strategies are not designed specifically for our dataset, they may not be universally applicable to all types of embedded firmware \update{and may result in a potential loss of precision}.
Targeting a specific firmware domain, analysts may be able to devise additional ranking strategies to further improve the accuracy of the fault localization. 
For instance, analysts can prioritize conditional instructions to verify the correctness of the control flow during the final manual investigation.

\subsection{Future Work}\label{sec:future-work}

While \system{} relies on an off-the-shelf fuzzer to collect the necessary runtime information, the data requirements could potentially be satisfied through other forms of dynamic program analysis as well. 
Therefore, our work is able to enhance other dynamic root cause analysis methods.

We acknowledge that \system{} encounters some analysis failures, which we believe could be mitigated by sophisticated ranking strategies and runtime information retrieval methods.
However, since they are incremental improvements upon \system{}, we leave them as future work.

Besides, \system{} currently analyzes code semantics at the granularity of instructions.
This leads to a limited understanding of the intentions of the whole program.
A promising future direction would be integrating \system{} with advanced techniques such as large language models (LLMs) to generate more comprehensive, natural language descriptions of the identified root causes. 

\section{Related Work}

\subsection{Embedded Firmware Fuzzing}

The embedded firmware runs on a variety of hardware platforms and interacts with multiple complicated peripherals.
Some early works conduct black-box fuzzing directly on real hardware devices \cite{koscher2010experimental,mulliner2011sms,chen2018iotfuzzer}, but this approach lacks comprehensive coverage guidance. 
$\mu$AFL \cite{li2022muafl} and SyzTrust \cite{wang2024syztrust} utilized a hardware debugger to collect runtime information with ETM feature.
Other works adopt hardware-in-the-loop \cite{zaddach2014avatar,koscher2015surrogates,muench2018avatar,talebi2018charm} approach to forward hardware requests to real peripherals.
However, these methods require frequent context switching and state synchronization between the emulator and hardware, leading to significant performance overhead.
Recent works adopt the rehosting technique for better scalability \cite{zheng2019firm,fasano2021sok,feng2020p2im,scharnowski2022fuzzware,scharnowski2023hoedur,zhou2021automatic,clements2020halucinator,chesser2023icicle}.
The key idea of rehosting is modeling peripherals and providing firmware with valid inputs, which mainly help firmware pass state checks during the initialization stage.
These rehosting works mainly focus on enabling fuzzing on IoT firmware, rather than tailoring and optimizing backend fuzzers (e.g., AFL \cite{afl}) for enhancing the fuzzing process effectiveness.

\subsection{Automated Fault Localization}

The manual investigation towards software crash analysis is a time-consuming and tedious process, which has motivated the development of automated fault localization techniques.

Spectrum-based approaches analyze crashing and non-crashing test cases to rank suspicious instructions based on their occurrence frequencies \cite{meyer2004comparison,jones2005empirical,abreu2006evaluation,abreu2007accuracy,naish2011model,wong2012software}. 
While effective, these methods are limited in their analysis granularity. 
More advanced techniques explore different execution paths near the initial crash and assign scores to program entities \cite{arumuga2007statistical,blazytko2020aurora,zhang2022default,liu2005sober,park2023benzene,xuracing}. 
However, the heavy time cost of analyzing each individual test case makes them impractical for large-scale firmware fuzzing.
Postmortem-based methods leverage post-crash artifacts, such as execution traces and memory dumps, to reversely analyze the program and identify the minimum set of instructions potentially responsible for the crash \cite{xu2017postmortem,mu2019pomp++,mu2019renn,wu2014crashlocator,cui2016retracer,cui2018rept,guo2019deepvsa}.
CrashLocator \cite{wu2014crashlocator} located faulty functions using the crash stack information in crash reports.
RETracer \cite{cui2016retracer} reversely analyzed the code and the stack to find out how a bad value propagates.
Besides, learning-based methods are also widely studied for a better understanding of crashes \cite{li2017transforming,sohn2017fluccs,li2019deepfl,li2021fault,zhang2023context,lou2021boosting,wang2021beep}. 
They proposed neural network architectures to analyze and incorporate the program context into suspicious score calculation.
Recently, with the rise of LLMs, program analysis has achieved vigorous advances \cite{yang2024large,shang2024far}.

\section{Conclusion}

In this paper, we present \system{}, a practical fault localization framework tailored specifically for embedded firmware.
\system{} leverages an event-based approach to precisely resolve the memory alias problem in postmortem-based fault localization methods.
We further propose a novel two-phase strategy to highlight instructions related to the crash and provide practical guidance for analysts in the final investigation.
We evaluate \system{} with \TotalTestCases{} crashing test cases across \TotalImages{} firmware images, achieving a success rate of \SystemSuccessRatio{}.
\CompareSummary{}

\section*{Acknowledgment}

We sincerely appreciate our shepherd and all the anonymous reviewers for their insightful and valuable feedback. 
This work was partly supported by the Fellowship of China National Postdoctoral Program for Innovative Talents (BX20230307), NSFC under Grant No. 62302443, Cisco Research Grant and Keysight Research Grant.

\bibliographystyle{IEEEtran}

\bibliography{sections/Ref}

\begin{thebibliography}{10}
\providecommand{\url}[1]{#1}
\csname url@samestyle\endcsname
\providecommand{\newblock}{\relax}
\providecommand{\bibinfo}[2]{#2}
\providecommand{\BIBentrySTDinterwordspacing}{\spaceskip=0pt\relax}
\providecommand{\BIBentryALTinterwordstretchfactor}{4}
\providecommand{\BIBentryALTinterwordspacing}{\spaceskip=\fontdimen2\font plus
\BIBentryALTinterwordstretchfactor\fontdimen3\font minus \fontdimen4\font\relax}
\providecommand{\BIBforeignlanguage}[2]{{%
\expandafter\ifx\csname l@#1\endcsname\relax
\typeout{** WARNING: IEEEtran.bst: No hyphenation pattern has been}%
\typeout{** loaded for the language `#1'. Using the pattern for}%
\typeout{** the default language instead.}%
\else
\language=\csname l@#1\endcsname
\fi
#2}}
\providecommand{\BIBdecl}{\relax}
\BIBdecl

\bibitem{wright2021challenges}
C.~Wright, W.~A. Moeglein, S.~Bagchi, M.~Kulkarni, and A.~A. Clements, ``Challenges in firmware re-hosting, emulation, and analysis,'' \emph{ACM Computing Surveys (CSUR)}, vol.~54, no.~1, pp. 1--36, 2021.

\bibitem{hpreport}
{HP}, ``{HP Wolf Security Threat Insights Report},'' \url{https://threatresearch.ext.hp.com/wp-content/uploads/2024/02/HP_Wolf_Security_Threat_Insights_Report_Q4_2023.pdf}, 2023, accessed Jun. 2024.

\bibitem{ibm-databreach}
{IBM}, ``{Cost of a Data Breach Report},'' \url{https://www.ibm.com/reports/data-breach}, 2023, accessed Jun. 2024.

\bibitem{viasat}
{Viasat}, ``{KA-SAT Network cyber attack overview},'' \url{https://news.viasat.com/blog/corporate/ka-sat-network-cyber-attack-overview}, 2022, accessed Jun. 2024.

\bibitem{scharnowski2022fuzzware}
T.~Scharnowski, N.~Bars, M.~Schloegel, E.~Gustafson, M.~Muench, G.~Vigna, C.~Kruegel, T.~Holz, and A.~Abbasi, ``Fuzzware: Using precise mmio modeling for effective firmware fuzzing,'' in \emph{31st USENIX Security Symposium (USENIX Security 22)}, 2022, pp. 1239--1256.

\bibitem{li2022muafl}
W.~Li, J.~Shi, F.~Li, J.~Lin, W.~Wang, and L.~Guan, ``$\mu$afl: non-intrusive feedback-driven fuzzing for microcontroller firmware,'' in \emph{Proceedings of the 44th International Conference on Software Engineering}, 2022, pp. 1--12.

\bibitem{zaddach2014avatar}
J.~Zaddach, L.~Bruno, A.~Francillon, D.~Balzarotti \emph{et~al.}, ``Avatar: A framework to support dynamic security analysis of embedded systems' firmwares.'' in \emph{NDSS}, vol.~14, 2014, pp. 1--16.

\bibitem{muench2018avatar}
M.~Muench, D.~Nisi, A.~Francillon, and D.~Balzarotti, ``Avatar 2: A multi-target orchestration platform,'' in \emph{Proc. Workshop Binary Anal. Res.(Colocated NDSS Symp.)}, vol.~18, 2018, pp. 1--11.

\bibitem{xu2017postmortem}
J.~Xu, D.~Mu, X.~Xing, P.~Liu, P.~Chen, and B.~Mao, ``Postmortem program analysis with hardware-enhanced post-crash artifacts,'' in \emph{26th USENIX Security Symposium (USENIX Security 17)}, 2017, pp. 17--32.

\bibitem{mu2022depth}
D.~Mu, Y.~Wu, Y.~Chen, Z.~Lin, C.~Yu, X.~Xing, and G.~Wang, ``An in-depth analysis of duplicated linux kernel bug reports.'' in \emph{NDSS}, 2022.

\bibitem{blazytko2020aurora}
T.~Blazytko, M.~Schl{\"o}gel, C.~Aschermann, A.~Abbasi, J.~Frank, S.~W{\"o}rner, and T.~Holz, ``{AURORA}: Statistical crash analysis for automated root cause explanation,'' in \emph{29th USENIX Security Symposium (USENIX Security 20)}, 2020, pp. 235--252.

\bibitem{park2023benzene}
Y.~Park, H.~Lee, J.~Jung, H.~Koo, and H.~K. Kim, ``Benzene: A practical root cause analysis system with an under-constrained state mutation,'' in \emph{2024 IEEE Symposium on Security and Privacy (SP)}.\hskip 1em plus 0.5em minus 0.4em\relax IEEE Computer Society, 2024, pp. 74--74.

\bibitem{xuracing}
D.~Xu, D.~Tang, Y.~Chen, X.~Wang, K.~Chen, H.~Tang, and L.~Li, ``Racing on the negative force: Efficient vulnerability root-cause analysis through reinforcement learning on counterexamples,'' 2024.

\bibitem{guo2019deepvsa}
W.~Guo, D.~Mu, X.~Xing, M.~Du, and D.~Song, ``{DEEPVSA}: Facilitating value-set analysis with deep learning for postmortem program analysis,'' in \emph{28th USENIX Security Symposium (USENIX Security 19)}, 2019, pp. 1787--1804.

\bibitem{cui2018rept}
W.~Cui, X.~Ge, B.~Kasikci, B.~Niu, U.~Sharma, R.~Wang, and I.~Yun, ``{REPT}: Reverse debugging of failures in deployed software,'' in \emph{13th USENIX Symposium on Operating Systems Design and Implementation (OSDI 18)}, 2018, pp. 17--32.

\bibitem{cui2016retracer}
W.~Cui, M.~Peinado, S.~K. Cha, Y.~Fratantonio, and V.~P. Kemerlis, ``Retracer: Triaging crashes by reverse execution from partial memory dumps,'' in \emph{Proceedings of the 38th International Conference on Software Engineering}, 2016, pp. 820--831.

\bibitem{li2017transforming}
X.~Li and L.~Zhang, ``Transforming programs and tests in tandem for fault localization,'' \emph{Proceedings of the ACM on Programming Languages}, vol.~1, no. OOPSLA, pp. 1--30, 2017.

\bibitem{sohn2017fluccs}
J.~Sohn and S.~Yoo, ``Fluccs: Using code and change metrics to improve fault localization,'' in \emph{Proceedings of the 26th ACM SIGSOFT International Symposium on Software Testing and Analysis}, 2017, pp. 273--283.

\bibitem{li2019deepfl}
X.~Li, W.~Li, Y.~Zhang, and L.~Zhang, ``Deepfl: Integrating multiple fault diagnosis dimensions for deep fault localization,'' in \emph{Proceedings of the 28th ACM SIGSOFT international symposium on software testing and analysis}, 2019, pp. 169--180.

\bibitem{yang2024large}
A.~Z. Yang, C.~Le~Goues, R.~Martins, and V.~Hellendoorn, ``Large language models for test-free fault localization,'' in \emph{Proceedings of the 46th IEEE/ACM International Conference on Software Engineering}, 2024, pp. 1--12.

\bibitem{wong2012software}
W.~E. Wong, V.~Debroy, Y.~Li, and R.~Gao, ``Software fault localization using dstar (d*),'' in \emph{2012 IEEE Sixth International Conference on Software Security and Reliability}.\hskip 1em plus 0.5em minus 0.4em\relax IEEE, 2012, pp. 21--30.

\bibitem{sarhan2022survey}
Q.~I. Sarhan and {\'A}.~Besz{\'e}des, ``A survey of challenges in spectrum-based software fault localization,'' \emph{IEEE Access}, vol.~10, pp. 10\,618--10\,639, 2022.

\bibitem{fasano2021sok}
A.~Fasano, T.~Ballo, M.~Muench, T.~Leek, A.~Bulekov, B.~Dolan-Gavitt, M.~Egele, A.~Francillon, L.~Lu, N.~Gregory \emph{et~al.}, ``Sok: Enabling security analyses of embedded systems via rehosting,'' in \emph{Proceedings of the 2021 ACM Asia conference on computer and communications security}, 2021, pp. 687--701.

\bibitem{scharnowski2023hoedur}
T.~Scharnowski, S.~W{\"o}rner, F.~Buchmann, N.~Bars, M.~Schloegel, and T.~Holz, ``Hoedur: embedded firmware fuzzing using multi-stream inputs,'' in \emph{32nd USENIX Security Symposium (USENIX Security 23)}, 2023, pp. 2885--2902.

\bibitem{chesser2023icicle}
M.~Chesser, S.~Nepal, and D.~C. Ranasinghe, ``Icicle: a re-designed emulator for grey-box firmware fuzzing,'' in \emph{Proceedings of the 32nd ACM SIGSOFT International Symposium on Software Testing and Analysis}, 2023, pp. 76--88.

\bibitem{unicorn}
\BIBentryALTinterwordspacing
{Unicorn Engine}, ``{The Ultimate CPU emulator},'' 2015, accessed Jun. 2024. [Online]. Available: \url{https://www.unicorn-engine.org/}
\BIBentrySTDinterwordspacing

\bibitem{alfred2007compilers}
V.~A. Alfred, S.~L. Monica, and D.~U. Jeffrey, \emph{Compilers Principles, Techniques \& Tools}.\hskip 1em plus 0.5em minus 0.4em\relax pearson Education, 2007.

\bibitem{mu2019pomp++}
D.~Mu, Y.~Du, J.~Xu, J.~Xu, X.~Xing, B.~Mao, and P.~Liu, ``{POMP++}: Facilitating postmortem program diagnosis with value-set analysis,'' \emph{IEEE Transactions on Software Engineering}, vol.~47, no.~9, pp. 1929--1942, 2019.

\bibitem{capstone}
{Capstone Engine}, ``{The Ultimate Disassembler},'' \url{https://www.capstone-engine.org/}, 2015, accessed Jun. 2024.

\bibitem{intelpt}
Intel, ``Collecting intel® processor trace (intel® pt) in intel® system debugger,'' \url{https://www.intel.com/content/www/us/en/developer/videos/collecting-processor-trace-in-intel-system-debugger.html?wapkw=intel%20pt}, accessed Jun. 2024.

\bibitem{fuzzware-dataset}
Fuzzware, ``Files used for reproducing fuzzware's experiments,'' \url{https://github.com/fuzzware-fuzzer/fuzzware-experiments/tree/main}, 2022, accessed Jun. 2024.

\bibitem{kochhar2016practitioners}
P.~S. Kochhar, X.~Xia, D.~Lo, and S.~Li, ``Practitioners' expectations on automated fault localization,'' in \emph{Proceedings of the 25th international symposium on software testing and analysis}, 2016, pp. 165--176.

\bibitem{etm}
{ARM}, ``Embedded trace macrocell, etmv1.0 to etmv3.5, architecture specification,'' \url{https://documentation-service.arm.com/static/5f90158b4966cd7c95fd5b5e}, 2011, accessed Jun. 2024.

\bibitem{zhang2023alligator}
Y.~Zhang, Y.~Hu, H.~Li, W.~Shi, Z.~Ning, X.~Luo, and F.~Zhang, ``Alligator in vest: A practical failure-diagnosis framework via arm hardware features,'' in \emph{Proceedings of the 32nd ACM SIGSOFT International Symposium on Software Testing and Analysis}, 2023, pp. 917--928.

\bibitem{koscher2010experimental}
K.~Koscher, A.~Czeskis, F.~Roesner, S.~Patel, T.~Kohno, S.~Checkoway, D.~McCoy, B.~Kantor, D.~Anderson, H.~Shacham \emph{et~al.}, ``Experimental security analysis of a modern automobile,'' in \emph{2010 IEEE symposium on security and privacy (SP)}.\hskip 1em plus 0.5em minus 0.4em\relax IEEE, 2010, pp. 447--462.

\bibitem{mulliner2011sms}
C.~Mulliner, N.~Golde, and J.-P. Seifert, ``{SMS} of death: From analyzing to attacking mobile phones on a large scale,'' in \emph{20th USENIX Security Symposium (USENIX Security 11)}, 2011.

\bibitem{chen2018iotfuzzer}
J.~Chen, W.~Diao, Q.~Zhao, C.~Zuo, Z.~Lin, X.~Wang, W.~C. Lau, M.~Sun, R.~Yang, and K.~Zhang, ``Iotfuzzer: Discovering memory corruptions in iot through app-based fuzzing.'' in \emph{NDSS}, 2018.

\bibitem{wang2024syztrust}
Q.~Wang, B.~Chang, S.~Ji, Y.~Tian, X.~Zhang, B.~Zhao, G.~Pan, C.~Lyu, M.~Payer, W.~Wang, and R.~Beyah, ``Syztrust: State-aware fuzzing on trusted os designed for iot devices,'' in \emph{2024 IEEE Symposium on Security and Privacy (SP)}.\hskip 1em plus 0.5em minus 0.4em\relax IEEE Computer Society, 2024, pp. 70--70.

\bibitem{koscher2015surrogates}
K.~Koscher, T.~Kohno, and D.~Molnar, ``{SURROGATES}: Enabling {Near-Real-Time} dynamic analyses of embedded systems,'' in \emph{9th USENIX Workshop on Offensive Technologies (WOOT 15)}, 2015.

\bibitem{talebi2018charm}
S.~M.~S. Talebi, H.~Tavakoli, H.~Zhang, Z.~Zhang, A.~A. Sani, and Z.~Qian, ``Charm: Facilitating dynamic analysis of device drivers of mobile systems,'' in \emph{27th USENIX Security Symposium (USENIX Security 18)}, 2018, pp. 291--307.

\bibitem{zheng2019firm}
Y.~Zheng, A.~Davanian, H.~Yin, C.~Song, H.~Zhu, and L.~Sun, ``{FIRM-AFL}:{High-Throughput} greybox fuzzing of {IoT} firmware via augmented process emulation,'' in \emph{28th USENIX Security Symposium (USENIX Security 19)}, 2019, pp. 1099--1114.

\bibitem{feng2020p2im}
B.~Feng, A.~Mera, and L.~Lu, ``{P2IM}: Scalable and hardware-independent firmware testing via automatic peripheral interface modeling,'' in \emph{29th USENIX Security Symposium (USENIX Security 20)}, 2020, pp. 1237--1254.

\bibitem{zhou2021automatic}
W.~Zhou, L.~Guan, P.~Liu, and Y.~Zhang, ``Automatic firmware emulation through invalidity-guided knowledge inference,'' in \emph{30th USENIX Security Symposium (USENIX Security 21)}, 2021, pp. 2007--2024.

\bibitem{clements2020halucinator}
A.~A. Clements, E.~Gustafson, T.~Scharnowski, P.~Grosen, D.~Fritz, C.~Kruegel, G.~Vigna, S.~Bagchi, and M.~Payer, ``{HALucinator}: Firmware re-hosting through abstraction layer emulation,'' in \emph{29th USENIX Security Symposium (USENIX Security 20)}, 2020, pp. 1201--1218.

\bibitem{afl}
{Michal Zalewski}, ``American fuzzy lop,'' \url{https://lcamtuf.coredump.cx/afl/}, 2010, accessed Jun. 2024.

\bibitem{meyer2004comparison}
A.~d.~S. Meyer, A.~A.~F. Garcia, A.~P.~d. Souza, and C.~L.~d. Souza~Jr, ``Comparison of similarity coefficients used for cluster analysis with dominant markers in maize (zea mays l),'' \emph{Genetics and Molecular Biology}, vol.~27, pp. 83--91, 2004.

\bibitem{jones2005empirical}
J.~A. Jones and M.~J. Harrold, ``Empirical evaluation of the tarantula automatic fault-localization technique,'' in \emph{Proceedings of the 20th IEEE/ACM international Conference on Automated software engineering}, 2005, pp. 273--282.

\bibitem{abreu2006evaluation}
R.~Abreu, P.~Zoeteweij, and A.~J. Van~Gemund, ``An evaluation of similarity coefficients for software fault localization,'' in \emph{2006 12th Pacific Rim International Symposium on Dependable Computing (PRDC'06)}.\hskip 1em plus 0.5em minus 0.4em\relax IEEE, 2006, pp. 39--46.

\bibitem{abreu2007accuracy}
------, ``On the accuracy of spectrum-based fault localization,'' in \emph{Testing: Academic and industrial conference practice and research techniques-MUTATION (TAICPART-MUTATION 2007)}.\hskip 1em plus 0.5em minus 0.4em\relax IEEE, 2007, pp. 89--98.

\bibitem{naish2011model}
L.~Naish, H.~J. Lee, and K.~Ramamohanarao, ``A model for spectra-based software diagnosis,'' \emph{ACM Transactions on software engineering and methodology (TOSEM)}, vol.~20, no.~3, pp. 1--32, 2011.

\bibitem{arumuga2007statistical}
P.~Arumuga~Nainar, T.~Chen, J.~Rosin, and B.~Liblit, ``Statistical debugging using compound boolean predicates,'' in \emph{Proceedings of the 2007 international symposium on Software testing and analysis}, 2007, pp. 5--15.

\bibitem{zhang2022default}
X.~Zhang, J.~Chen, C.~Feng, R.~Li, W.~Diao, K.~Zhang, J.~Lei, and C.~Tang, ``Default: mutual information-based crash triage for massive crashes,'' in \emph{Proceedings of the 44th International Conference on Software Engineering}, 2022, pp. 635--646.

\bibitem{liu2005sober}
C.~Liu, X.~Yan, L.~Fei, J.~Han, and S.~P. Midkiff, ``Sober: statistical model-based bug localization,'' \emph{ACM SIGSOFT Software Engineering Notes}, vol.~30, no.~5, pp. 286--295, 2005.

\bibitem{mu2019renn}
D.~Mu, W.~Guo, A.~Cuevas, Y.~Chen, J.~Gai, X.~Xing, B.~Mao, and C.~Song, ``Renn: Efficient reverse execution with neural-network-assisted alias analysis,'' in \emph{2019 34th IEEE/ACM International Conference on Automated Software Engineering (ASE)}.\hskip 1em plus 0.5em minus 0.4em\relax IEEE, 2019, pp. 924--935.

\bibitem{wu2014crashlocator}
R.~Wu, H.~Zhang, S.-C. Cheung, and S.~Kim, ``Crashlocator: Locating crashing faults based on crash stacks,'' in \emph{Proceedings of the 2014 International Symposium on Software Testing and Analysis}, 2014, pp. 204--214.

\bibitem{li2021fault}
Y.~Li, S.~Wang, and T.~Nguyen, ``Fault localization with code coverage representation learning,'' in \emph{2021 IEEE/ACM 43rd International Conference on Software Engineering (ICSE)}.\hskip 1em plus 0.5em minus 0.4em\relax IEEE, 2021, pp. 661--673.

\bibitem{zhang2023context}
Z.~Zhang, Y.~Lei, X.~Mao, M.~Yan, X.~Xia, and D.~Lo, ``Context-aware neural fault localization,'' \emph{IEEE Transactions on Software Engineering}, 2023.

\bibitem{lou2021boosting}
Y.~Lou, Q.~Zhu, J.~Dong, X.~Li, Z.~Sun, D.~Hao, L.~Zhang, and L.~Zhang, ``Boosting coverage-based fault localization via graph-based representation learning,'' in \emph{Proceedings of the 29th ACM Joint Meeting on European Software Engineering Conference and Symposium on the Foundations of Software Engineering}, 2021, pp. 664--676.

\bibitem{wang2021beep}
S.~Wang, K.~Liu, B.~Lin, L.~Li, J.~Klein, X.~Mao, and T.~F. Bissyand{\'e}, ``Beep: Fine-grained fix localization by learning to predict buggy code elements,'' \emph{arXiv preprint arXiv:2111.07739}, 2021.

\bibitem{shang2024far}
X.~Shang, S.~Cheng, G.~Chen, Y.~Zhang, L.~Hu, X.~Yu, G.~Li, W.~Zhang, and N.~Yu, ``How far have we gone in stripped binary code understanding using large language models,'' \emph{arXiv preprint arXiv:2404.09836}, 2024.

\end{thebibliography}

\newpage
%\appendix % ACM template
\appendices % IEEE template

\section{Dataset}\label{app:crash-id}
We follow the official guidance of Fuzzware to deploy the framework in its reproducing mode, and finally successfully reproduce \TotalTestCases{} test cases that exhibit crashing behaviors.
These test cases encompass a total of \TotalImages{} firmware images across \TotalPlatforms{} hardware platforms.
One firmware sample can be compiled from the same source code and deployed across multiple platforms.
As shown in Table~\ref{tbl:evaluation-dataset}, we assign crash IDs from $C_1$ to $C_{\TotalTestCases{}}$ for these test cases for simplicity.
The dataset includes both synthetic ($C_1$-$C_{9}$)  and real-world ($C_{10}$-$C_{41}$) test cases.

\begin{table}[htbp]
    \centering
    \caption{Hardware platforms, firmware samples, and direct crash reasons corresponding to crash IDs in \system{}'s evaluation.}
    \resizebox{0.95\linewidth}{!}{
    \begin{tabular}{@{}lllc@{}}
    \toprule
    ID & Hardware Platform & Firmware Sample & Direct Crash Reason            \\ \midrule
    $C_{1}$  & ARCH\_PRO         & Pw\_Discovery                & Invalid Memory Read \\
    $C_{2}$  & EFM32GG\_STK3700  & Pw\_Discovery                & Invalid Memory Read \\
    $C_{3}$  & EFM32LG\_STK3600  & Pw\_Discovery                & Invalid Memory Read \\
    $C_{4}$  & LPC1549           & Pw\_Discovery                & Invalid Memory Read \\
    $C_{5}$  & LPC1768           & Pw\_Discovery                & Invalid Memory Read \\
    $C_{6}$  & MOTE\_L152RC      & Pw\_Discovery                & Invalid Memory Read \\
    $C_{7}$  & NUCLEO\_F103RB    & Pw\_Discovery                & Invalid Memory Read \\
    $C_{8}$  & NUCLEO\_L152RE    & Pw\_Discovery                & Invalid Memory Read \\
    $C_{9}$  & UBLOX\_C027       & Pw\_Discovery                & Invalid Memory Read \\
    $C_{10}$ & STM32F429ZI       & CNC                          & Invalid Instruction Execution \\
    $C_{11}$ & STM32F103RB       & Gateway                      & Invalid Memory Read \\
    $C_{12}$ & STM32F429ZI       & PLC                          & Invalid Memory Read \\
    $C_{13}$ & STM32F429ZI       & PLC                          & Invalid Memory Read \\
    $C_{14}$ & STM32F103RB       & Robot                        & Invalid Memory Read \\
    $C_{15}$ & STM32F103RB       & Gateway                      & Invalid Memory Read \\
    $C_{16}$ & STM32F103RB       & Gateway                      & Invalid Memory Write \\
    $C_{17}$ & STM32F103RB       & Gateway                      & Invalid Memory Read \\
    $C_{18}$ & STM32F103RB       & Gateway                      & Invalid Memory Read \\
    $C_{19}$ & STM32F429ZI       & PLC                          & Invalid Memory Read \\
    $C_{20}$ & STM32F429ZI       & utasker\_USB                 & Invalid Memory Read \\
    $C_{21}$ & MAX32600          & Thermostat                   & Invalid Instruction Execution \\
    $C_{22}$ & MAX32600          & RF\_Door\_Lock               & Invalid Instruction Execution \\
    $C_{23}$ & SAMR21            & 6LoWPAN\_Receiver            & Invalid Memory Write \\
    $C_{24}$ & SAMR21            & 6LoWPAN\_Sender              & Invalid Memory Write \\
    $C_{25}$ & MAX32600          & RF\_Door\_Lock               & Invalid Memory Write \\
    $C_{26}$ & STM32F103RE       & 3DPrinter                    & Invalid Instruction Execution \\
    $C_{27}$ & STM32F429ZI       & utasker\_MODBUS              & Invalid Memory Read \\
    $C_{28}$ & STM32F429ZI       & utasker\_MODBUS              & Invalid Memory Read \\
    $C_{29}$ & STM32L432KC       & Zepyhr\_SocketCan            & Invalid Memory Read \\
    $C_{30}$ & STM32L432KC       & Zepyhr\_SocketCan            & Invalid Memory Read \\
    $C_{31}$ & STM32F429ZI       & utasker\_USB                 & Invalid Instruction Execution \\
    $C_{32}$ & SAM4E\_XPRO       & Socket\_Echo\_Server         & Invalid Instruction Execution \\
    $C_{33}$ & DISCO\_L475\_IOT1 & Bluetooth\_Peripheral\_HIDs  & Invalid Memory Read \\
    $C_{34}$ & DISCO\_L475\_IOT1 & Bluetooth\_Peripheral\_HIDs  & Invalid Memory Read \\
    $C_{35}$ & SAM4S\_XPLAINED   & Socket\_Echo\_Server         & Invalid Memory Read \\
    $C_{36}$ & SAM4S\_XPLAINED   & Socket\_Echo\_Server         & Invalid Memory Read \\
    $C_{37}$ & NRF52840          & Bluetooth\_Peripheral\_HIDs  & Invalid Memory Write \\
    $C_{38}$ & SAM4S\_XPLAINED   & Bluetooth\_Peripheral\_HIDs  & Invalid Memory Write\\
    $C_{39}$ & CC2538            & hello-world                  & Invalid Memory Read\\
    $C_{40}$ & CC2538            & snmp-server                  & Invalid Memory Read\\
    $C_{41}$ & CC2538            & hello-world                  & Invalid Memory Write\\ \bottomrule
    \end{tabular}
    }
    \label{tbl:evaluation-dataset}
\end{table}

\begin{figure*}[thpb]
    \centering
    \includegraphics[width=1\textwidth]{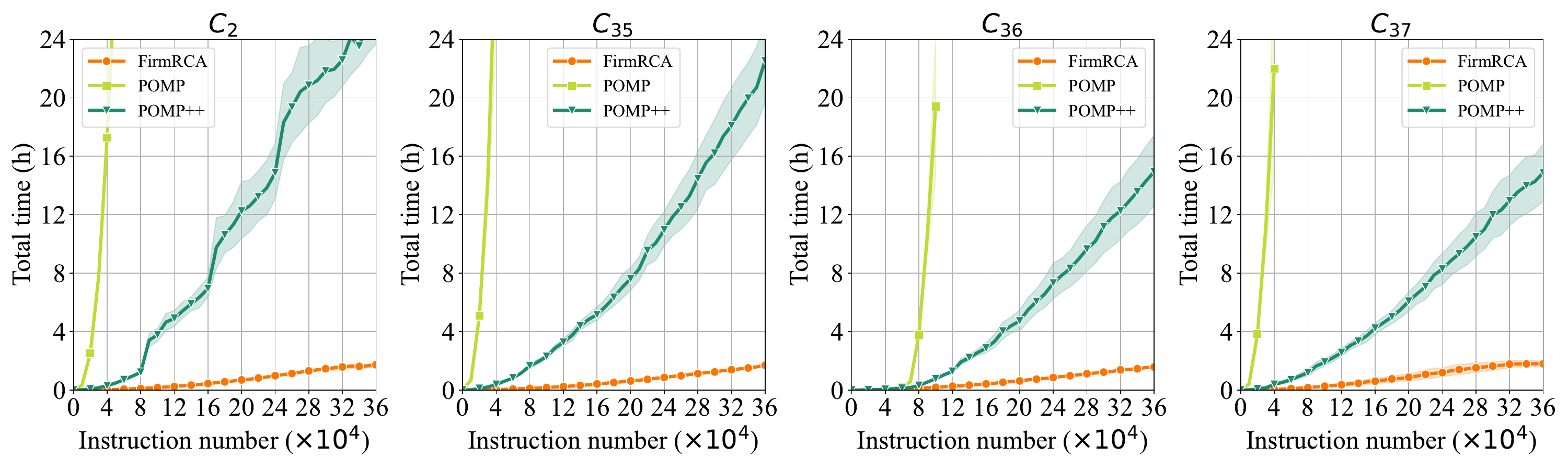} 
    \caption{Overall time costs of fault localization on the different number of instructions under analysis ($C_{2}$, $C_{35}$, $C_{36}$, and $C_{37}$).}
    \label{fig:multiple-time-growth} 
\end{figure*}

\section{Footprint Collection Overhead}\label{app:hook-overhead}
We reproduce each test case 50 times, both with and without the data and action event, to measure the overhead of our event-based footprint collection approach and eliminate randomness. 
As shown in Table \ref{tbl:events-overhead}, our event-based approach only introduces a small amount of extra overhead, with an average time of \AvgOverheadTimes{} and an average log file size of \AvgOverheadSpace{}.

\begin{table}[htbp]
    \centering
    \caption{Time and space overhead of reproducing a crashing test case. $\mathrm{T}$ denotes time cost and $\mathrm{S}$ denotes the size of event files. The subscript $\mathrm{o}$, $\mathrm{d}$, $\mathrm{a}$, and $\mathrm{w}$ denote no events enabled, data events enabled, action events enabled, and both events enabled, respectively. $\mathrm{S_o}$ is omitted (all values are 0MB) because no files are generated with both events disabled.}
    \resizebox{1\linewidth}{!}{
    \begin{tabular}{@{}crrrrrrr@{}}
    \toprule
    {ID}  & $\mathrm{T_o}$ (s) & $\mathrm{T_d}$ (s) & $\mathrm{T_a}$ (s) & $\mathrm{T_w}$ (s) & $\mathrm{S_d}$ (MB) & $\mathrm{S_a}$ (MB) & $\mathrm{S_w}$ (MB) \\ \midrule
    $C_{1}$  & 0.77 & 1.18 & 1.56 & 1.98 & 1.60 & 1.99 & 3.58  \\
    $C_{2}$  & 0.74 & 2.69 & 4.05 & 5.72 & 11.74 & 14.18 & 25.92  \\
    $C_{3}$  & 0.74 & 5.82 & 8.83 & 13.53 & 31.97 & 38.28 & 70.25  \\
    $C_{4}$  & 0.79 & 1.95 & 2.85 & 3.74 & 6.89 & 8.70 & 15.59  \\
    $C_{5}$  & 0.71 & 1.34 & 1.62 & 1.77 & 1.08 & 1.38 & 2.47  \\
    $C_{6}$  & 0.84 & 1.52 & 2.03 & 2.31 & 3.27 & 4.13 & 7.40  \\
    $C_{7}$  & 0.69 & 4.22 & 6.20 & 9.16 & 19.73 & 23.65 & 43.38  \\
    $C_{8}$  & 0.77 & 1.35 & 1.64 & 2.17 & 2.00 & 2.63 & 4.63  \\
    $C_{9}$  & 0.62 & 3.62 & 5.30 & 7.88 & 16.88 & 20.02 & 36.90  \\
    $C_{10}$  & 0.80 & 1.65 & 2.13 & 2.42 & 2.87 & 4.27 & 7.09  \\
    $C_{11}$  & 0.92 & 1.42 & 2.55 & 2.88 & 2.51 & 6.43 & 8.94  \\
    $C_{12}$  & 0.66 & 1.48 & 2.99 & 3.56 & 2.64 & 8.99 & 11.63  \\
    $C_{13}$  & 0.78 & 1.94 & 2.43 & 3.18 & 5.39 & 6.13 & 11.52  \\
    $C_{14}$  & 0.92 & 1.57 & 1.59 & 1.72 & 0.53 & 0.79 & 1.31  \\
    $C_{15}$  & 0.63 & 1.07 & 1.29 & 1.05 & 0.26 & 0.46 & 0.71  \\
    $C_{16}$  & 0.69 & 1.82 & 2.96 & 3.33 & 3.35 & 8.31 & 11.51  \\
    $C_{17}$  & 0.87 & 1.58 & 2.81 & 3.10 & 2.56 & 6.63 & 9.19  \\
    $C_{18}$  & 0.91 & 1.41 & 2.45 & 2.99 & 2.48 & 6.47 & 8.98  \\
    $C_{19}$  & 0.73 & 1.05 & 1.49 & 1.27 & 0.17 & 0.27 & 0.44  \\
    $C_{20}$  & 0.89 & 2.06 & 5.48 & 6.40 & 5.77 & 19.98 & 25.76  \\
    $C_{21}$  & 0.76 & 1.68 & 2.33 & 2.78 & 3.64 & 5.76 & 9.40  \\
    $C_{22}$  & 0.80 & 2.08 & 3.00 & 3.73 & 6.30 & 9.10 & 15.39  \\
    $C_{23}$  & 1.33 & 1.85 & 2.20 & 2.60 & 2.15 & 3.40 & 5.55  \\
    $C_{24}$  & 0.84 & 1.51 & 1.78 & 1.78 & 1.38 & 2.02 & 3.32  \\
    $C_{25}$  & 0.57 & 2.65 & 4.30 & 5.94 & 9.85 & 16.56 & 26.41  \\
    $C_{26}$  & 0.65 & 1.47 & 2.72 & 2.96 & 2.27 & 7.49 & 9.75  \\
    $C_{27}$  & 0.76 & 1.40 & 1.40 & 1.72 & 1.01 & 1.80 & 2.80  \\
    $C_{28}$  & 0.95 & 1.25 & 1.60 & 1.66 & 1.07 & 1.86 & 2.94  \\
    $C_{29}$  & 0.73 & 1.69 & 2.13 & 2.47 & 2.19 & 3.21 & 5.40  \\
    $C_{30}$  & 0.79 & 1.26 & 1.89 & 2.36 & 0.88 & 3.60 & 6.07  \\
    $C_{31}$  & 0.74 & 1.91 & 5.36 & 5.94 & 5.61 & 19.64 & 25.24  \\
    $C_{32}$  & 0.98 & 5.90 & 9.24 & 13.52 & 29.23 & 37.95 & 67.11  \\
    $C_{33}$  & 1.01 & 2.15 & 2.98 & 3.81 & 5.02 & 7.96 & 12.98  \\
    $C_{34}$  & 0.82 & 2.04 & 2.93 & 3.28 & 4.63 & 6.14 & 10.77  \\
    $C_{35}$  & 0.98 & 9.63 & 16.47 & 24.88 & 57.54 & 74.78 & 132.28  \\
    $C_{36}$  & 1.02 & 5.16 & 9.31 & 13.41 & 27.40 & 38.76 & 66.14  \\
    $C_{37}$  & 1.18 & 6.61 & 12.84 & 18.03 & 34.81 & 56.24 & 91.06  \\
    $C_{38}$  & 0.81 & 7.14 & 11.30 & 17.22 & 38.54 & 49.61 & 88.11  \\
    $C_{39}$  & 0.74 & 1.80 & 2.25 & 2.41 & 2.67 & 3.00 & 5.68  \\
    $C_{40}$  & 0.80 & 1.61 & 1.84 & 2.21 & 1.22 & 1.99 & 3.21  \\
    $C_{41}$  & 0.93 & 6.26 & 6.92 & 11.32 & 28.70 & 23.84 & 52.56  \\
    \midrule
    Avg. & 0.82 & 2.63 & 4.07 & 5.47 & 9.51 & 13.62 & 23.16 \\
    \bottomrule
    \end{tabular}
    }
    \label{tbl:events-overhead}
    \end{table}

\section{Implementation of Comparative Works}\label{app:migration}

We select POMP and POMP++ for comparison because they are the latest open-sourced postmortem-based fault localization works.
We carefully inspect the source code of POMP and POMP++ and migrate them to the 32-bit ARM Cortem-M architecture in the following aspects.

$\bullet$ \textbf{Runtime information collection.}
POMP and POMP++ rely on hardware feature support (i.e., Intel PT) to obtain the complete execution trace.
In order to provide the same level of information, we examine the data structure used by POMP and POMP++ and modify our event-based footprint collection method to offer data in their formats.

$\bullet$ \textbf{Reverse execution.}
POMP and POMP++ target x86 software, which is totally different from firmware in terms of the instruction set architecture.
We design inverse handlers for the 32-bit ARM Cortex-M architecture for the inverse execution of the reverse execution module.
In our comparison, POMP, POMP++, and \system{} share the same inverse handlers to prevent extra overhead and side effects.

$\bullet$ \textbf{Value-set analysis.}
POMP++ employs a customized value-set analysis and leverages static information (e.g., memory region) to solve the memory alias problem.
We also migrate their method to the 32-bit ARM Cortex-M architecture to identify memory regions for a fair comparison.
Note that POMP++ also utilizes heuristics that only exist in the x86 architecture, such as a special function \instruction{get\_pc\_thunk.cx}, to improve the region identification.
We ignore such heuristics because they do not apply in firmware scenarios.

We acknowledge that \system{} and POMP/POMP++ do not share exactly the same scenario, which may introduce performance differences.
We argue that since \system{} is focused on firmware fault localization as a new post-fuzzing work, our event-based method is competitive with their runtime information collection (e.g., full execution trace collection through Intel PT), exhibiting great effectiveness and efficiency in fault localization.

\section{The Growth of Overall Time Costs}\label{app:multiple-time-growth}
Figure~\ref{fig:multiple-time-growth} shows four crashing test cases with more than 360,000 instructions in their execution traces.
We repeatedly perform fault localization five times with \system{}, POMP, and POMP++ with a timeout of 24 hours.

% \newpage % The Meta-Review should at least start on a new column

% Use \appendices and not \appendix due to IEEEtran.cls quirks
% \appendices % if not used earlier

\section{Meta-Review}

The following meta-review was prepared by the program committee for the 2025
IEEE Symposium on Security and Privacy (S\&P) as part of the review process as
detailed in the call for papers.

\subsection{Summary}

\system{} presents a fault localization system for ARM-based firmware. It uses dynamic analysis to identify the root cause of crashes. It leverages event-based footprint collection and data propagation tracking to resolve challenges like memory aliasing. \system{} operates in two phases: first, it collects runtime data to trace memory accesses and instructions, and second, it applies reverse execution and ranking heuristics to identify and rank instructions leading to the crash. Evaluated on 41 crash cases across 17 firmware images, \system{} achieves a 92.7\% success rate in root cause identification and outperforms similar methods.

\subsection{Scientific Contributions}
\begin{itemize}
\item Independent Confirmation of Important Results with Limited Prior Research
\item Creates a New Tool to Enable Future Science
\item Addresses a Long-Known Issue
\item Provides a Valuable Step Forward in an Established Field
\end{itemize}

\subsection{Reasons for Acceptance}
\begin{enumerate}
\item The paper addresses the issue of improving fault localization techniques in ARM-based firmware, an underexplored area.
\item It introduces an event-based analysis method and ranking heuristic that significantly improves root cause identification accuracy, outperforming existing tools like POMP and POMP++.
\item The \system{} method performs well (a 92.7\% success rate on crash analysis).
\end{enumerate}

\subsection{Noteworthy Concerns} % Exclude if your meta-review does not have noteworthy concerns
\begin{enumerate} % Enumerate environment is not necessary if there is only one
\item Several reviewers noted concerns about the evaluation being restricted to open-source targets, giving the paper an advantage.
\item Some reviewers questioned the novelty of the approach (e.g., reverse execution is not new, and what authors are suggesting is to just adapt it for embedded systems).
\item Concerns regarding reproducibility were raised due to the reliance on reverse execution, which has known challenges, and the fact that the paper does not discusses these issues.
\end{enumerate}

\section{Response to the Meta-Review} % Optional

\textbf{Response to 1.}
The meta-review notes that the evaluation being restricted to open-source targets.
To clarify, \system{} chooses raw binaries as targets for the best applicability.
These binaries are all closed-source, in line with the common assumption that the firmware is usually closed-source.

\textbf{Response to 2.}
The meta-review notes that the novelty of the approach is questioned.
We argue that the automated firmware crash analysis remains an underexplored area. Previous works fail to analyze deep root causes with extremely long execution traces due to several unresolved challenges in the firmware scenario, including inadequate debugging mechanisms and limited investigation guidance.
Therefore, we need novel designs to address these challenges for firmware crash analysis. 
Specifically, we propose an event-based method to selectively collect memory accesses and executed instructions of crashing test cases. 
\system{} addresses the memory alias problem, enabling deep root cause analysis with an extremely long execution trace.
Furthermore, we point out that the over-tainted instructions provided by previous postmortem-based works are impractical, and propose a ranking mechanism to offer practical guidance for the final manual investigation.

\textbf{Response to 3.}
The meta-review notes that reproducibility concerns are raised due to the reliance on reverse execution, which has known challenges.
In \system{}, the reverse execution is performed after crash reproduction, so it does not introduce a reproduction issue. 
Additionally, \system{} concentrates on reproducible crashes, which is consistent with recent RCA research focus.

% Dear Shepherd,

% Thank you for your comments! We appreciate your help and would like to accept the meta-review (Option 1).

% According to the reviewers’ comments, we plan to revise our paper, specifically adjusting the discussion of our introduction examples, figure descriptions, failed cases analysis and limitations in the paper. 
% Please let us know if/what additional comments you have and we'd be happy to revise the paper accordingly.

% Thank you for your time and consideration.

% Best,
% Authors

% Dear Shepherd,

% Thank you for your comments! We appreciate your help and would like to accept the meta-review with no changes (Option 1).

% Best, Authors

% Dear Shepherd,

% Thank you for your comments again! We would like to accept the meta-review with no changes (Option 1).

% Our near-final camera-ready version has been uploaded through the "Edit Submission" form. Besides, we also attach the same version along with the comment. Please review our paper at your convenience. Thanks!

% Best, Authors

% Dear Shepherd,

% Thank you for your comments! We have updated the meta-review according to your latest comment.

% The new near-final camera-ready version has also been uploaded through the "Edit Submission" form. Please check our paper at your convenience. Thanks!

% Best, Authors

\end{document}